\documentstyle[preprint,aps,epsf,floats,tighten]{revtex}
\begin{document}
\draft
\preprint{\vbox{\hbox{UTPT-99-14}}}
\title{Symmetry breaking via fermion 4-point functions}
\author{ F. S. Roux\thanks{roux@physics.utoronto.ca},
T.~Torma\thanks{kakukk@physics.utoronto.ca} and
B. Holdom\thanks{bob.holdom@utoronto.ca}}
\address{Department of Physics\\University of Toronto\\Toronto, Ontario\\Canada M5S~1A7}
\date{\today}
\maketitle
\begin{abstract} 
We construct the effective action and
gap equations for nonperturbative fermion 4-point functions. Our results apply to
situations in which fermion masses can be ignored, which is the case for theories of
strong flavor interactions involving standard quarks and leptons above the electroweak
scale. The structure of the gap equations is different from what a naive generalization of the
2-point case would suggest, and we find for example that gauge exchanges
are insufficient to generate nonperturbative 4-point functions when the number of colors is
large.
\end{abstract}

\tableofcontents


\section{Introduction}

In theories without elementary scalar fields, electroweak symmetry breaking is 
accomplished through the dynamical generation of fermion masses of order a TeV. This
chiral symmetry breaking must somehow be fed down to the lighter quark and lepton
masses to produce their mass spectrum. The original ``extended technicolor'' idea was
to accomplish this via broken gauge interactions, with symmetry breaking scales
ranging up to $\sim 1000$~TeV. There are two main problems with this idea. One is that
it immediately raises the question of what breaks these new interactions. What plays
the role of the order parameter for this breakdown? If it is another mass-type order
parameter then it must involve new fermions with exotic quantum numbers, since
conventional quarks and leptons are not allowed to have mass above the electroweak
breaking scale. The other problem is that the complicated structure of the quark and
lepton mass spectrum is directly reflected in the structure of the new gauge
interactions. This led to models having an input structure (in the form of the symmetry
and representation content) no less complicated than the output structure (the quark
and lepton mass spectrum), and basically the whole extended technicolor idea gradually
sank under its own weight.

Our motivation for the current work comes from the well know fact that there is a much richer
set of possible order parameters which may be utilized by a strongly interacting theory, beyond
the simple mass-type order parameter. Therefore the order parameters that signal the
breakdown of flavor symmetries could actually be constructed out of the known quarks and
leptons, due to their participation in strong flavor dynamics at high scales. The only constraint
is that these order parameters respect electroweak symmetry, which in turn ensures that quarks
and leptons do not develop masses on these scales and thus remain in the theory below the
flavor scale. Among such $SU(2)_L\times U(1)$ invariant order parameters are ones involving
four fermions. They will result in effective 4-fermion interactions in the low energy theory, and
thus the same order parameters responsible for breaking flavor symmetries can be responsible
for feeding down the TeV masses to lighter quark and leptons. The difference from extended
technicolor models is that we now have a larger class of possible operators, and the variety in
size and structure of these operators must be a reflection of strong dynamics, rather than a
reflection of some input structure.

We mention in passing that there are 2-fermion order parameters involving standard quarks
and leptons, of the chirality preserving type, which could serve as flavor breaking order
parameters. If right-handed neutrinos are appended to
the standard model set of fermions, then a Majorana-mass-type order parameter could also be
considered. In this work we shall restrict ourselves to the study of the more diverse set of
4-fermion order parameters, which we shall refer to as (nonperturbative) 4-point functions.

We are entertaining the possibility that strong interactions dynamically break various
flavor symmetries, but preserve $SU(2)_L\times U(1)$. For this to happen the gauge
interactions may have to be chiral, so that the gauge interactions themselves resist
the formation of masses. For example a competition between different gauge
interactions could end up reducing the attraction in the mass channels. In nonchiral
theories as well, it could simply be that the
critical coupling required for the formation of a more symmetric 4-point function is
less than that for any 2-point function. In any case we stress that {\it any} strongly
interacting theory of flavor above the electroweak scale must preserve $SU(2)_L\times
U(1)$.
With the neglect of fermion masses the chirality structure of the various 4-point
functions will be the focus of our attention, and we often keep the flavor and gauge
structures implicit. The chirality structure is constrained by $SU(2)_L\times U(1)$
symmetry and in particular 4-point functions having an odd number of each chirality,
such as $\langle \overline{\psi} \partial\!\!\!/ \psi \psi \overline{\psi} \rangle$
are excluded.

To illustrate this discussion consider a strong $SU(N_c)$ gauge interaction for an even
number $N_f>2$ of massless fermions in the fundamental representation of the gauge
group. The flavor symmetry is $SU(N_f)_L \times SU(N_f)_R \times U(1)_V$. The
chirality preserving 4-point functions can be made invariant under this chiral
symmetry, while the chirality changing 4-point functions cannot be invariant. The
formation of the latter can still be consistent with a chiral isospin (or
`electroweak') symmetry and a vector `family' symmetry,
\begin{equation}
SU(2)_L \times SU(2)_R \times SU(N_f/2)_V \times U(1)_V .
\label{f2}
\end{equation}
The corresponding chirality changing 4-point functions have the structure
\begin{equation}
\left\langle \left( \overline{\psi}_R^{Aa} \psi_L^{Ab} \right) \epsilon_{ac}
\epsilon_{bd} \left( \overline{\psi}_R^{Bc} \psi_L^{Bd} \right) \right\rangle,
\label{f3}
\end{equation}
where the upper case superscripts denote family, the lower case superscripts denote
isospin and $\epsilon_{ac}$ is a $2 \times 2$ anti-symmetric matrix.

An investigation into the nonperturbative generation of 4-point functions requires
knowledge of the effective action and gap equations of the 4-point functions. We use
the procedures in \cite{r_dm} to derive these tools in a general context. Before we
present the detailed derivations, we provide an overview which summarizes the
derivations. We will also explore the solutions to the gap equations in the framework
of the model of the previous paragraph, in the large $N_c$ limit.

\section{Overview of the derivations}
\label{over}

More than three decades ago De Dominicis and Martin published a set of papers
\cite{r_dm} in which they provide procedures for the derivation of gap equations and
effective actions for various $n$-point Green functions.  Although their analysis was
presented in the framework of non-relativistic statistical physics the procedures can
be readily applied to quantum field theory.\footnote{This was already shown many years
ago by Cornwell, Jackiw and Tomboulis\cite{r_cjt} for the 2-point case. Although they
used a different procedure to obtain it, their famous effective action is precisely
the quantum field theory generalization of the De Dominicis-Martin effective action
for 2-point functions.} We follow their procedure for 4-point functions to derive the
effective action and gap equations in the context of gauge theories with massless
fermions.

Below we review their procedure in its adapted form. We are only interested in the part
of their analysis dealing with 4-point functions. There are two parts to this
procedure, which can be regarded as two independent ways to derive the required gap
equations.  One part provides a direct derivation of these equations.  The other part
is a derivation of the effective action, from which one can then reproduce the same
gap equations by imposing a stationarity condition of the form
\begin{equation}
{\delta \Gamma \over \delta C} = 0,
\label{o3a}
\end{equation}
where $C$ denotes a 4-point function. Although the direct method is perhaps a quicker
method to obtain the expressions for the gap equations, it is the other method which
actually makes a statement about the vacuum of the theory by showing that these gap
equations extremize the effective action.

\begin{figure}
\centerline{\epsfysize = 2 cm \epsfbox{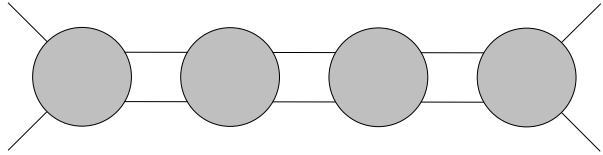}}
\caption{A bubble chain}
\label{chain}
\end{figure}

The direct method is presented in detail in Sec.~\ref{gap}. The goal is to derive the gap
equations for the five 4-point functions as presented
in (\ref{v6}). Schematically they are all of the form
\begin{equation}
C = T_s(C) + T_d(C) + T_d^T(C) + K(C,g) 
\label{o3b}
\end{equation}
where $C$ is a connected, amputated 4-point function. The $T$'s are sums of bubble chains,
shown in Fig.~\ref{chain}, formed by connecting $C$'s through pairs of fermion lines. There
are three different bubble chains because there are three ways to pair off the
four external lines of a 4-point function. The subscript $s$ ($d$) indicates that the fermion lines
in each pair have the same (opposite) direction. The
superscript $T$ indicates that the two incoming or two outgoing lines are interchanged, with
`incoming' and `outgoing' referring to the arrows on the external fermion lines. Only the
$K$-term depends on the gauge coupling explicitly, and it contains all the 4-particle irreducible
(4PI) diagrams\footnote{In short, a diagram that cannot be separated into two parts by cutting
four fermion lines is 4-particle irreducible. See Sec.~\ref{4part} for a better
description.} that one can form from these $C$'s together with the other propagators and
vertices of the gauge theory. The diagrams are also 2-particle irreducible (2PI) since the
fermion line represents the full fermion propagator. We are assuming that the fermion
propagator itself does not break any symmetries and in particular is massless.

The main task in the direct derivation of the gap equations is to find the expressions
for the $T$'s in terms of the $C$'s; here we sketch the procedure.
We require the following definition:\ when a 4-point diagram cannot be separated into
two parts where each part has a pair of the original external lines, by cutting two
internal fermion lines, the 4-point diagram is said to be {\it simple} with respect to
this particular pairing of external lines.  The sum of all 4-point diagrams that are
simple with respect to one specific pairing of external lines is denoted by $U$. One
can define $T$ as the sum of all 4-point diagrams that are simple with respect to
{\it all but} the aforementioned pairing of external lines. (A 4-point function can
only be nonsimple with respect to one pairing, which follows from the absence of
fermion 3-point functions.) Then $C=U+T$ for this pairing of external lines. Since
there are three ways of pairing off the four external lines one can define three $U$'s
and three $T$'s.
\begin{figure}
\centerline{\epsfysize = 2 cm \epsfbox{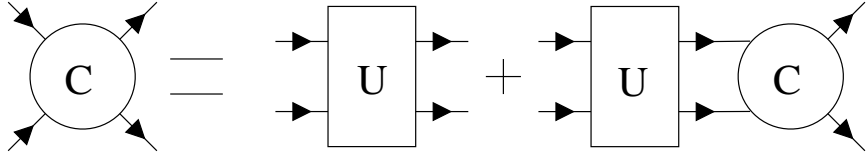}}
\caption{Master equation}
\label{master}
\end{figure}
Next, a {\it master equation} relates $C$ to one of the three
$U$'s, for a particular pairing of external lines. The equation, shown
diagrammatically in Fig.~\ref{master}, can be written as
\begin{equation}
C = U + U C .
\label{o5}
\end{equation}
The 2-line connection of the $U$ to the $C$ is presented
as a product of two 4-point functions in (\ref{o5}). This notation will be used
throughout the rest of this paper, and it will be discussed further in
Sec.~\ref{algebra}.  From (\ref{o5}) it directly follows that 
\begin{equation}
U = C [1 + C]^{-1} ,
\label{o6}
\end{equation}
and thus the $T$-term in the gap equation for a
particular pairing is given  by
\begin{equation}
T = C - U = C^2 [1 + C]^{-1} .
\label{o7}
\end{equation}
The actual derivations of the various $T$'s which appear in the five gap
equations are more involved. The different chiralities lead to sets of more
complicated master equations which are to be solved simultaneously to obtain the
expressions for the $U$'s and the $T$'s. The expressions of these $T$'s are provided
in (\ref{ap5}), (\ref{ap25}), (\ref{ap26}) and (\ref{ap27}).

As discussed in Sec.~\ref{4part} the
$K$-term is simple with respect to all pairings. It can be generated from
the sum of all 4PI vacuum diagrams through the following functional derivation
\begin{equation}
K[C] = \left[{\delta V_{4PI}[C] \over \delta C}\right]_{amp} .
\label{o9}
\end{equation}
Here $V_{4PI}[C]$ denotes the sum of all 4PI vacuum diagrams in which the 4-point
function, $C$, is treated as a 4-point vertex. The subscript $amp$ indicates that the
diagrams are amputated. The $K$-term together with the expressions for the
$T$'s completes the derivation of the gap equations in (\ref{v6}).

\begin{figure}
\centerline{\epsfysize = 5 cm \epsfbox{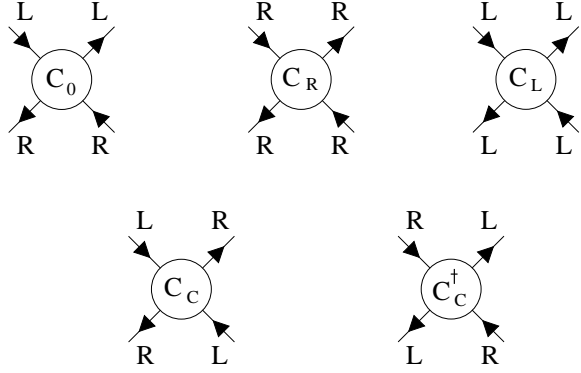}}
\caption{The five connected, amputated 4-point functions}
\label{c's}
\end{figure}

Next we present an overview of the derivation of the effective action from which one
can derive the same gap equations. The resulting expression for the effective action
as a functional of the various $C$'s is
\begin{eqnarray}
\Gamma & = & {\rm Tr} \left \{ \ln \left(1 - \frac{1}{4} R_s C^{\dag} L_s C \right)_s
+ \ln \left(1 + \frac{1}{2} C_R \right)_s + \ln \left(1 + \frac{1}{2} C_L \right)_s +
\ln(1+C_0)_s \right.  \nonumber \\ & & + \frac{1}{2} \ln \left(1 - R C_0 L C_0
\right)_d + \frac{1}{2} \ln \left(1 + C_R \right)_d + \frac{1}{2} \ln \left(1 +
C_L \right)_d + \ln \left(1 + C_0 \right)_d \nonumber \\ & & + \frac{1}{2} \ln
\left(1 - Z C^{\dag} Z C \right)_d - C_R - C_L - 2 C_0 \nonumber \\ & & \left.+
\frac{1}{2} C C^{\dag} + C_0^2 + \frac{1}{4} C_R^2 + \frac{1}{4} C_L^2 \right \}
+ V_{4PI}[C,C^{\dag},C_0,C_R,C_L] .
\label{o10}
\end{eqnarray}
The five different $C$'s which appear in this expression are defined as follows (see
Fig.~\ref{c's}). When all four of the external lines of a 4-point function have the
same chirality, say left-handed (right-handed), we denote the 4-point function by
$C_L$ ($C_R$).  These two 4-point functions are chirality preserving.  The
other chirality preserving 4-point function, which we denote by $C_0$, has one
incoming and one outgoing line of each chirality.  The remaining two 4-point functions
are chirality changing; their incoming and outgoing lines have the opposite chirality. 
When the chirality of their incoming lines are left-handed (right-handed) we denote
them by $C_C$ ($C_C^{\dag}$). The other symbols which appear in this expression are
defined as follows
\begin{mathletters}
\label{ap1}
\begin{eqnarray}
R_s & = & \left[1+\frac{1}{2} C_R \right]_s^{-1} \label{ap1rs} \\
L_s & = & \left[1+\frac{1}{2} C_L \right]_s^{-1} \label{ap1ls} \\
R & = & \left[1+ C_R \right]_d^{-1} \label{ap1r} \\
L & = & \left[1+ C_L \right]_d^{-1} \label{ap1l} \\
Z & = & \left[1+C_0 \right]_d^{-1} \label{ap1z} ,
\end{eqnarray}
\end{mathletters}
where the subscript $s$ and $d$ have the same meaning as in (\ref{o3b}). The
formulation of these objects and how they are applied in the derivation of the
effective action are discussed in Sec.~\ref{algebra}.

The effective action is the Legendre transform of the sum of all connected vacuum
diagrams in the presence of the 4-fermion sources. The $J$-dependence is replaced by a
$C$-dependence, and thus the aim is to find an expression for the effective action in
terms of the 4-point functions ($C$'s) with only implicit dependence on the sources
($J$'s). The problem is to avoid overcounting. This
problem is solved in \cite{r_dm} by adding and subtracting various sets of diagrams in such a
way that the overcounting is eliminated and the original sum of vacuum diagrams is retained.
This is done with the aid of the following topological equation, which is proved in \cite{r_dm}.
\begin{equation}
1 =  N_{skel} + N_{vert} - N_{art} + \sum_{pair ~ cycles} [1 - N_p + N_{pp}] 
\label{o11}
\end{equation}
Each term in this equation denotes the number of a specific element or feature (such as
skeletons or pair cycles) which is present in a vacuum diagram. These terms and the
way they are applied to vacuum diagrams are explained in
Section~\ref{sec:topol},\ref{paircycles}. The topological relation relates the numbers of times
a diagram is overcounted when considering the various manifestations of the various features.
It is useful because, by distinguishing a specific feature of the vacuum diagrams, one can
generate the associated term from the $C$'s. In other words, the topological equation will
allows us to express the sum of all connected vacuum diagrams in terms of various sets
of vacuum diagrams which can be explicitly constructed from the 4-point functions.

By performing functional derivatives with respect to the various $C$'s on the
effective action in (\ref{o10}), one can reproduce the exact expressions for the gap
equations.  This is shown in Sec.~\ref{derive}. The expressions for the effective
action and the gap equations represent the main result of our analysis, which extends
to 4-point functions the CJT analysis \cite{r_cjt} of 2-point functions. The structure of our
equations are different from what a naive extension of the CJT analysis would suggest, and
this is made evident in the following application.

\section{Gap equations in the large $N_c$ limit}

The above discussion summarizes the program carried out in the rest of the paper. In
this section we will make a first attempt to use the resulting formalism to extract
information about the 4-point functions. We will treat the model described in the
introduction, a $SU(N_c)$ gauge theory with $N_f$ fermions, each having $N_c$ colors. 
Although such a theory is not chiral, we can still look for any intrinsic tendency of
the theory to produce nonperturbative 4-point functions. We consider a large
$N_c$ expansion to simplify the structure of the gap equations. We then need to know
the leading $N_c$-dependences of the $C$'s.  Note that the gap equations contain
sequences of 4-point functions of the form $[ 1 + C ]^{-1}$. In these sequences
adjacent $C$'s are interconnected with two fermion lines, which can form color loops,
and each color loop gives a factor of $N_c$. In the large $N_c$ limit these sequences
would only be convergent and nontrivial if the $C$'s themselves are $O(1/N_c)$. This
is true for all types of $C$'s.

\begin{figure}
\centerline{\epsfysize = 2 cm \epsfbox{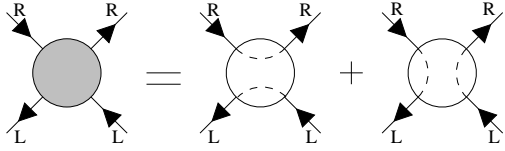}}
\caption{Color structures of $C_0$.}
\label{color}
\end{figure}

The next step is to separate each of the different $C$'s into {\it color structures}.
If the color indices on the external lines of $C$ are $a,b,\overline{c}$ and
$\overline{d}$, then one can connect these indices internally in the following two
ways: $\delta^{a\overline{c}} \delta^{b\overline{d}}$ or $\delta^{a\overline{d}}
\delta^{b\overline{c}}$. $C_0$ does not have any symmetry with respect to interchanges of
external lines and so in this case the two color structures represent two different objects,
\begin{equation}
C_0 = C_{0V} + C_{0S} .
\label{o14}
\end{equation}
These color structures are shown in Fig.~\ref{color}, where the dashed lines indicate
how the color indices are connected internally. The connected indices have
the same chirality for $C_{0V}$ and the opposite chirality for $C_{0S}$. 
The two color structures for the other $C$'s are just transposed versions
of the same object. By `transposed' we mean that the two lines with the outgoing
fermion arrows are interchanged, or equivalently the two incoming lines are
interchanged. We denote a particular color structure of these $C$'s
with a hat, and to reconstruct the original $C$ we write
\begin{equation}
C = \hat{C} + (\hat{C})^T .
\label{o15}
\end{equation}

\begin{figure}
\centerline{\epsfysize = 3 cm \epsfbox{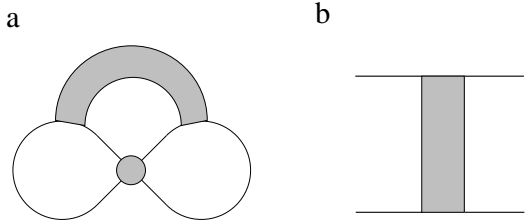}}
\caption{The 4PI gauge interaction vacuum and 4-point diagrams at leading order in
$1/N_c$. }
\label{kterm}
\end{figure}

The $1/N_c$ expansion is obtained by substituting (\ref{o14}) and (\ref{o15}) into the
various expressions, paying attention to the color loops formed by the various color
structure orientations of the $C$'s.
\begin{figure}
\centerline{\epsfysize = 3 cm \epsfbox{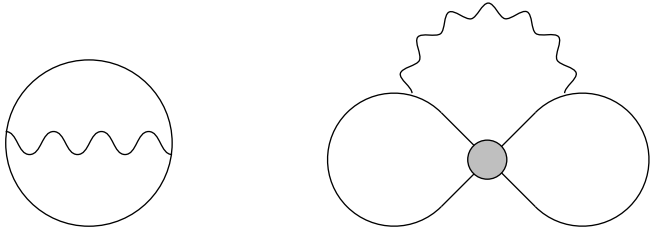}}
\caption{The two 4PI vacuum diagrams with one-gauge-boson exchanges}
\label{ogbe}
\end{figure}
The great simplification that occurs is due to the fact that the diagrams in $V_{4PI}$ at leading
order in $1/N_c$ contain no more than one 4-point function each. The two vacuum diagrams
that are leading order in $1/N_c$, as well as being leading order in $\alpha$, are shown in
Fig.~\ref{ogbe}.\footnote{We note that these are the only 4PI diagrams at leading order in the
gauge coupling, to all orders in $1/N_c$.} Because the fermions are massless the single 4-point
function that appears in the second diagram in Fig.~\ref{ogbe} is chirality preserving.  All the
other diagrams in $V_{4PI}$
involving a 4-point function at leading order in $1/N_c$ also look similar to this diagram. The
single gauge boson is just replaced by the set of 4PI planar graphs. We denote
the sum of all these vacuum diagrams by Fig.~\ref{kterm}a and the 4-point diagram
that is obtained after removing the 4-point function by Fig.~\ref{kterm}b.

The expression for the effective action at leading order in $1/N_c$ splits into two parts.
These two parts decouple from each other because they do not share the same $C$'s.
One part is
\begin{equation}
\Gamma_1 = {\rm Tr} \left \{ \frac{1}{2} \ln \left( 1 - \hat{C}_C \hat{Z}
\hat{C}_C^{\dag} \hat{Z} \right)_d + \ln \left( 1 + C_{0S} \right)_d -  C_{0S} \right
\} + V_{4PI,1}
\label{nc1}
\end{equation}
where $\hat{Z}=[1+C_{0S}]_d^{-1}$ and $V_{4PI,1}$ contains the diagram in
Fig.~\ref{kterm}a with $C_{0S}$ as the 4-point function. The other part is
\begin{eqnarray}
\Gamma_2 &=& {\rm Tr} \left \{ \frac{1}{2} \ln \left( 1 - C_{0V} \hat{L} C_{0V}
\hat{R} \right)_d + \frac{1}{2} \ln \left( 1 + \hat{C}_R \right)_d + \frac{1}{2} \ln
\left( 1 + \hat{C}_L \right)_d - \frac{1}{2} \hat{C}_R - \frac{1}{2} \hat{C}_L \right
\}
\nonumber\\
&& + V_{4PI,2}
\label{nc2}
\end{eqnarray}
where $\hat{R} = [1 + \hat{C}_R]_d^{-1}$, $\hat{L} = [1 + \hat{C}_L]_d^{-1}$ and
$V_{4PI,2}$ contains the diagram in Fig.~\ref{kterm}a with $C_L$ or $C_R$ as the
4-point function. The gap equations associated with (\ref{nc1}) are
\begin{equation}
\hat{C}_C = T_d , ~~~ \hat{C}_C^{\dag} = T^{\dag}_d ~~~ {\rm and} ~~~ C_{0S} =
T^0_d + K_0
\label{nc3}
\end{equation}
and those for (\ref{nc2}) are
\begin{equation}
\hat{C}_R = T^R_d + K_R , ~~~ \hat{C}_L = T^L_d + K_L ~~~ {\rm and} ~~~ C_{0V} =
T^{\prime}_d .
\label{nc4}
\end{equation}
The various $T$'s are provided in (\ref{ap5}), (\ref{ap25}), (\ref{ap26}) and
(\ref{ap27}), and the various $K$'s appear in (\ref{v6}). The gap equations for
$\hat{C}_C$ and $\hat{C}_C^{\dag}$ do not have $K$-terms, because we have seen that in
the large
$N_c$ limit the 4PI vacuum diagrams are independent of $\hat{C}_C$ and
$\hat{C}_C^{\dag}$. The expression for $C_{0V}$ in (\ref{nc4}) also does not have a
$K$-term, because the color structure of the would-be $K$-term does not match. It shows
up in the $C_{0S}$-expression in (\ref{nc3}) instead.

Consider the first set of gap equations in
(\ref{nc3}), where we look first at $\hat{C}_C$ (or equivalently $\hat{C}_C^{\dag}$).
From the expression for
$T_d$, given in (\ref{ap27a}), we have:
\begin{equation}
\hat{C}_C = \hat{C}_C - \hat{Z} \hat{C}_C \hat{Z} \left[ 1 - \hat{C}_C^{\dag} \hat{Z}
\hat{C}_C \hat{Z} \right]_d^{-1} 
\label{nc5}
\end{equation}
where $\hat{Z} = [1 + C_{0S}]_d^{-1}$. It is clear that the only solution for this
expression is $\hat{C}_C=0$, irrespective of what the value of $C_{0S}$ is. This
happens because the equation does not have a $K$-term.

Using the expression for
$T^0_d$, (\ref{ap27c}), the gap equation for $C_{0S}$ becomes:
\begin{equation}
C_{0S} = C_{0S} + \hat{Z} \left[ 1 - \hat{C}_C^{\dag} \hat{Z} \hat{C}_C \hat{Z}
\right]_d^{-1} - 1 + X_0 .
\label{nc6}
\end{equation}
Here $X_0$, which is a contribution from the $K$-term, denotes the 4PI planar graphs in
Fig.~\ref{kterm}b, with the chiralities of the external fermion lines the same as $C_0$. For
$\hat{C}_C=\hat{C}_C^{\dag}=0$, (\ref{nc6}) gives
\begin{equation}
C_{0S} = X_0 \left[ 1 - X_0 \right]_d^{-1} ,
\label{nc6b}
\end{equation}
which implies that $C_{0S}$ is generated perturbatively by the sum of `ladder'
diagrams, with each rung the set of 4PI planar graphs.

A 4-point function of the $C_{0S}$ form may have a nontrivial flavor structure
which would prevent it from being generated perturbatively, i.e.~its gap equation would not
have a $K$-term:
\begin{equation}
C_{0S} = \hat{C}_C^{\dag} \hat{Z} \hat{C}_C .
\label{nc7}
\end{equation}
In this case it is only possible to generate a nonperturbative $C_{0S}$ indirectly
through $\hat{C}_C$ and $\hat{C}_C^{\dag}$, if the latter were nonzero.

Consider next the other set of gap equations in (\ref{nc4}). Using the expression
for
$T^{\prime}_d$, (\ref{ap26a}), the gap equation for $C_{0V}$ becomes
\begin{equation}
C_{0V} = C_{0V} - \hat{R} C_{0V} \hat{L} \left[ 1 - C_{0V} \hat{R} C_{0V} \hat{L}
\right]_d^{-1} 
\label{nc8}
\end{equation}
where $\hat{R} = [1 + \hat{C}_R]_d^{-1}$ and $\hat{L} = [1 + \hat{C}_L]_d^{-1}$. This
is very similar to (\ref{nc5}), and again the only solution is $C_{0V}=0$. As for $\hat{C}_R$
(or $\hat{C}_L$), by using the expression for $T^R_d$ (\ref{ap26c}) we obtain
\begin{equation}
\hat{C}_R = \hat{C}_R + \hat{R} \left[ 1 - C_{0V} \hat{L} C_{0V} \hat{R}
\right]_d^{-1} - 1 + X_R.
\label{nc9}
\end{equation}
Here $X_R$ denotes the 4PI planar graphs in Fig.~\ref{kterm}b where the chiralities of
the external fermion lines are those of a $C_R$. This gives
\begin{equation}
\hat{C}_R = X_R \left[ 1 - X_R \right]_d^{-1} 
\label{nc10}
\end{equation}
since $C_{0V}=0$. Hence, just like $C_{0S}$, $\hat{C}_R$ is generated perturbatively
by a `ladder' sum of 4PI planar graphs.

\begin{figure}
\centerline{\epsfysize = 3 cm \epsfbox{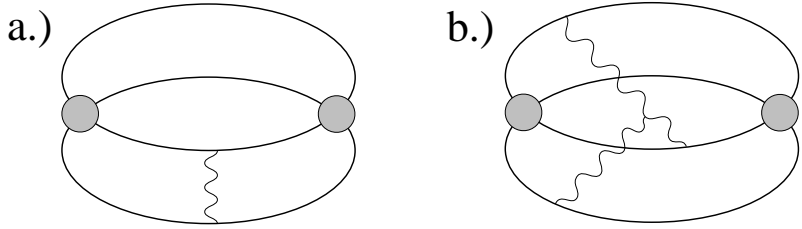}}
\caption{Vacuum diagrams with two 4-point functions. }
\label{trip}
\end{figure}

The absence of
nonperturbative solutions of gap equations for 4-point functions seems counterintuitive in the
light of the situation which is so well known for 2-point functions\cite{r_p}. The naive
generalization of the linearized ladder gap equation for the fermion self-energy to 4-point
functions would be of the form
\begin{equation}
C_C = E C_C 
\label{dis3}
\end{equation}
where $E$ is the one gauge exchange kernel. Equations of this type were considered in
\cite{r_ht,r_mw}.\footnote{An effective infrared cutoff, which is necessary for nontrivial
solutions and which appears naturally in the 2-point case, had to be postulated in
\cite{r_ht,r_mw}.} The reason we do not obtain such an equation can be traced to the
absence in
$V_{4PI}$ of the diagram consisting of two 4-point functions and one-gauge-boson
exchange, shown in Fig.~\ref{trip}a. Its absence is independent of the $1/N_c$ expansion
and is due simply to the fact that this diagram is not 4PI. This diagram is implicitly included
in the
$T$-terms in (\ref{o3b}) (since the chirality preserving 4-point functions implicitly have
one-gauge-boson exchange contributions), and we have seen that the $T$-terms by
themselves cannot generate nonperturbative solutions. Note that the diagram involving a
triple gauge vertex, shown in Fig.~\ref{trip}b, {\it is} included in $V_{4PI}$, but this
contribution is subleading in $1/N_c$.

In light of these results it is of interest to consider the possible effects of
instantons, which lie outside of our current approach. Due to the chirality changing
nature of instanton effects, they may lead to the dynamical generation of the
chirality changing 4-point functions, $\hat{C}_C$ and $\hat{C}_C^{\dag}$. We note that
the lines of a 't Hooft-vertex operator\cite{r_t} can be closed off with chirality
changing 4-point functions, and that the diagram containing chirality changing 4-point
functions only (no chirality preserving 4-point functions) would be leading in
$1/N_c$. Thus it may be the case that at leading order in $1/N_c$, $\hat{C}_C$ and
$\hat{C}_C^{\dag}$ are generated dynamically by instanton effects. $C_{0S}$ would then
be affected as in (\ref{nc7}).

We are finding then a qualitative difference between the 2-point function case and the
4-point function case, in the large $N_c$ limit. None of the 4-point functions can be
generated dynamically by gauge exchanges, and only the chirality changing 4-point
functions, $C_C$ and $C_C^{\dag}$, could in principle be dynamically generated via
instanton effects. We remind the reader that these particular 4-point functions are
relevant to the symmetry breaking pattern in (\ref{f2}), which is of special interest for
models of flavor physics which preserve electroweak symmetries.

We now present the detailed derivation of the gap equations and the effective action.
First we present the general formulation and notation in Sec.~\ref{green}. The direct
derivation of the gap equations is presented in Sec.~\ref{gap} while Sec.~\ref{effek}
contains the derivation of the effective action. In Sec.~\ref{derive} we show that the
gap equations minimize the effective action. 

\section{Green functions and 4-point functions}
\label{green}

In order to extend the CJT analysis \cite{r_cjt} to treat 4-point functions, one can introduce a
nonlocal 4-fermion source term in addition to the nonlocal 2-fermion source term. The
generating functional is then given by,
\begin{equation}
W^{\prime}[j,J] = -i\ln \left[ \int \exp \left(i S + i S_j + i S_J \right) {\cal D} \right] ,
\label{ei6a}
\end{equation}
where $S$ is the action of the gauge theory and ${\cal D}$ is the functional measure
over all the fields in the theory. Schematically
\begin{eqnarray}
S_j & = & \int \overline\psi(x_1) j(x_1,x_2) \psi(x_2)\ d^4x_1 d^4x_2 \\ S_J & = & \int
\overline\psi(x_1) \psi(x_2) J(x_1 ...  x_4) \overline\psi(x_3) \psi(x_4)\ d^4x_1 ... 
d^4x_4 . \nonumber
\end{eqnarray}
Here $j$ denotes the nonlocal 2-fermion source and $J$ denotes the nonlocal 4-fermion
source. The generating functional, $W^{\prime}[j,J]$, is the sum of all connected vacuum
diagrams which one can construct by using the Feynman rules of a gauge theory,
together with the sources as 2-fermion and 4-fermion vertices.

We perform two Legendre transforms on this generating functional to arrive at the
effective action. The first Legendre transformation,
\begin{equation}
\Gamma_{CJT}[S,J] = W^{\prime}[j,J]-\int j\cdot S ,
\end{equation}
replaces the functional dependences $j$ by a functional dependences on the full
fermion propagator $S$, which is expressed as
\begin{equation}
S=\frac{\delta W^{\prime}}{\delta j} .
\end{equation}
$\Gamma_{CJT}[S,J]$ is the CJT effective action \cite{r_cjt} in the presence of four-fermion
sources. The stationarity condition
\begin{equation}
\frac{\delta\Gamma_{CJT}[S,J]}{\delta S}=0,
\end{equation}
provides a way to determine the full propagator $S$ up to a dependence on the 4-fermion
source.

At this point it would be opportune
to discuss the chiral properties of the
Green functions. There are five nonlocal 4-fermion source terms which are for this
purpose included in the action, and so the 4-fermion source term in (\ref{ei6a})
becomes
\begin{eqnarray}
S_J & = & \int \left( \frac{1}{4} \overline{\psi}_R(x_1) \psi_L(x_2) J_C(x_1 ...  x_4)
\overline{\psi}_R(x_3) \psi_L(x_4) \right. \nonumber \\ & & + \frac{1}{4}
\overline{\psi}_L(x_1) \psi_R(x_2) J_C^{\dag}(x_1 ...  x_4) \overline{\psi}_L(x_3)
\psi_R(x_4) \nonumber \\ & & + \overline{\psi}_R(x_1) \psi_L(x_2) J_0(x_1 ...  x_4)
\overline{\psi}_L(x_3) \psi_R(x_4) \nonumber \\ & & + \frac{1}{4} \overline{\psi}_R(x_1)
\psi_R(x_2) J_R(x_1 ...  x_4) \overline{\psi}_R(x_3) \psi_R(x_4) \nonumber \\ & & \left. +
\frac{1}{4} \overline{\psi}_L(x_1) \psi_L(x_2) J_L(x_1 ...  x_4) \overline{\psi}_L(x_3)
\psi_L(x_4) \right)\  d^4x_1 ...  d^4x_4 .
\label{ei6}
\end{eqnarray}
The sources with the subscript $C$ are associated with chirality changing Green
functions, while the other three sources are chirality preserving.  Four of the source terms
have the same chiralities on both $\psi$-fields, as well as on both $\overline{\psi}$-fields. 
Thus the symmetry associated with the identical incoming pair and the identical outgoing pair
introduces the symmetry factor of $\frac{1}{4}$. The color and flavor indices of the fermion
fields are contracted on the sources.

$\Gamma_{CJT}[S,J]$ can be expressed as the sum of one-loop terms, $-{\rm Tr}
\{ \ln(S^{-1})
\} + {\rm Tr} \{(S^{-1}-\partial\!\!\!/)S)\}$, which are independent of the 4-fermion sources,
and $W[S,J]$, the set of connected 2PI vacuum diagrams with fermion lines representing the
full propagator.
From now on we will treat the
$W[S,J]$ as our generating functional.

The second Legendre transform gives the effective action of interest,
\begin{equation}
\Gamma^{\prime}[S,G] = W[S,J] - \int \left( \frac{1}{4} J_C G_C^{\dag} + \frac{1}{4}
J_C^{\dag} G_C + J_0 G_0 + \frac{1}{4} J_R G_R + \frac{1}{4}J_L G_L \right)\ d^4x_1
...  d^4x_4 .
\label{ei7}
\end{equation}
The Green functions are given by the functional derivatives of the generating functional
with respect to the appropriate sources,
\begin{equation}
{\delta W \over \delta J_C^{\dag}} = G_C , ~~~
{\delta W \over \delta J_C} = G_C^{\dag} , ~~~
{\delta W \over \delta J_0} = G_0 , ~~~
{\delta W \over \delta J_R} = G_R ~~~ {\rm and} ~~~
{\delta W \over \delta J_L} = G_L .
\label{ei8}
\end{equation}
Both the sources and the Green functions have color and flavor indices. The effective action
obeys the following stationarity conditions
\begin{equation}
{\delta \Gamma^{\prime}[S,G] \over \delta G_C^{\dag}} = 0 , ~~~
{\delta \Gamma^{\prime}[S,G] \over \delta G_C} = 0 , ~~~
{\delta \Gamma^{\prime}[S,G] \over \delta G_0} = 0 , ~~~
{\delta \Gamma^{\prime}[S,G] \over \delta G_R} = 0 ~~~ {\rm and} ~~~
{\delta \Gamma^{\prime}[S,G] \over \delta G_L} = 0 .
\label{ei8a}
\end{equation}

Now we see that in spite of the fact that the addition of a nonlocal 4-fermion source term
would change the full propagator, it is consistent to fix $S$ and vary only the Green
function $G$. Throughout our discussions we assume a general massless propagator,
\begin{equation} 
S(p) = i { Z(p) \over p\!\!\!/}.
\label{ei11} 
\end{equation}
It is also more convenient to write the effective action as a functional of the connected
and amputated parts of the Green functions. These objects are denoted by the symbol $C$
and we shall refer to them as {\it 4-point functions} to distinguish them from the
Green functions,
\begin{equation} 
C(x_1...x_4) = \left[ \langle T \overline{\psi}_1 \psi_2 \overline{\psi}_3 \psi_4 \rangle -
\langle T \overline{\psi}_1 \psi_2 \rangle \langle T \overline{\psi}_3 \psi_4 \rangle -
\langle T \overline{\psi}_1 \psi_4 \rangle \langle T \overline{\psi}_3 \psi_2 \rangle
\right]_{amp} ,
\label{ei9} 
\end{equation}
where the Green function is
\begin{equation} 
G = \langle T \overline{\psi}_1 \psi_2 \overline{\psi}_3 \psi_4 \rangle .
\label{ei10} 
\end{equation}
The 4-point functions, $C$'s, are classified in the same way as the Green functions
(recall Fig.~\ref{c's} and the discussion below (\ref{o10})). Therefore in
place of $\Gamma^{\prime}[S,G]$ we consider $\Gamma[C]$, in terms of which the
stationarity conditions become
\begin{equation}
{\delta \Gamma[C] \over \delta C_C^{\dag}} = 0 , ~~~
{\delta \Gamma[C] \over \delta C_C} = 0 , ~~~
{\delta \Gamma[C] \over \delta C_0} = 0 , ~~~
{\delta \Gamma[C] \over \delta C_R} = 0 ~~~ {\rm and} ~~~
{\delta \Gamma[C] \over \delta C_L} = 0 .
\label{ei12}
\end{equation}
These conditions are equivalent to (\ref{ei8a}) because the full propagators are held
constant. The stationarity of the effective action with respect to the 4-point
functions ($C$'s) contains the nonperturbative information we are trying
to extract.

The derivation of the effective action involves finding a diagrammatic
representation in which the five terms that are removed through the Legendre
transformation are the only ones with explicit sources. All other source terms should
end up inside subdiagrams which are replaced by explicit $C$'s, so that we have a
formula for $\Gamma$ which contains no reference to $J$.

\section{Direct derivation of the gap equations}
\label{gap}

The gap equations we seek will be expressed in terms of 2PI diagrams constructed out of
the 4-point functions (treated as a 4-point vertex), 2-point functions (our full
fermion propagator) and the vertices and propagators associated with a gauge
theory. We do not need the 4-fermion sources for the derivation in this
section, they will appear again in Sec.~\ref{effek}. Some
diagrams will contain gauge interactions explicitly while others will not, and it is
by balancing these two sets of diagrams that one can find a nontrivial solution for
the 4-point functions under investigation.  The infinite set of diagrams with explicit
gauge interaction cannot, of course, be given in closed form, while a set of the
diagrams consisting only of 4-point functions and fermion propagators can be. The
latter set are the bubble chains (see Fig.~\ref{chain}) which consist of sequences of
4-point functions.

\subsection{Preliminaries}
\label{algebra}

\subsubsection{Types of fermion pairs}

We first describe our notation. All the 4-point functions which are considered here
(that is, all the $C$'s, $U$'s and $T$'s, which are defined below) are amputated.
Therefore, when two of these 4-point functions are connected by two fermion lines, two
fermion propagators are to be inserted between the two 4-point functions.  We shall
leave this step implicit in our notation, and so $C_C C_C^{\dag}$ really means $C_C S S
C_C^{\dag}$.  Since $S$ is chirality preserving the product of two 4-point functions is
only possible if the directions and chiralities of the fermion lines that are to be
connected are compatible.  This may depend on the type of 4-point function (a $C_C$ can
never by directly connected to another $C_C$) or on the type of fermion pairs.

\begin{figure} 
\centerline{\epsfysize = 8 cm \epsfbox{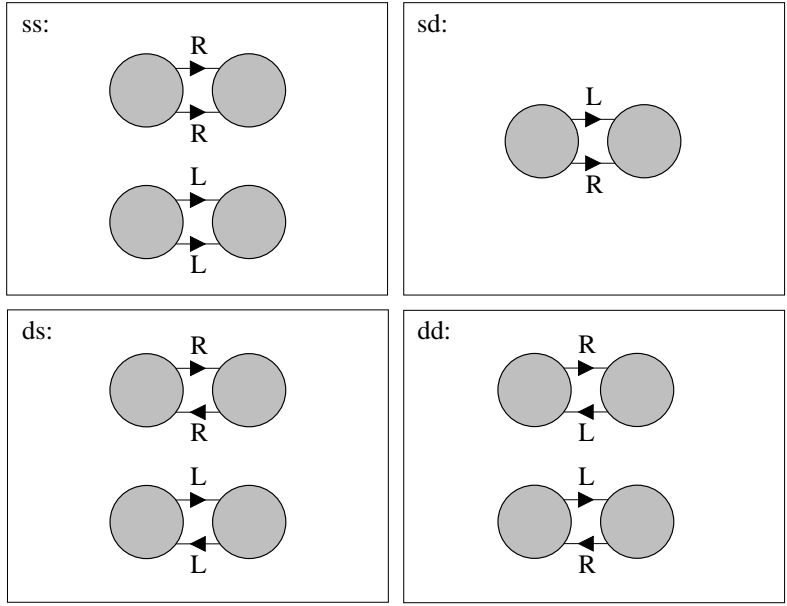}}
\caption{Four types of fermion pairs that can connect 4-point functions.}
\label{pairs}
\end{figure}

We define four types of fermion pairs depending on the directions and chiralities of
the fermion lines which connect two adjacent 4-point functions.  Each type of fermion
pair is denoted by two letters, the first of which is associated with the directions
and the second with the chiralities of the fermion lines.  We use {\it s} and {\it d}
to denote `same' and `different' respectively.
$ss$, $sd$, $ds$, or
$dd$ will appear as subscripts on objects or brackets to indicate that the objects or the enclosed
expressions have a specific type of fermion pairs, as illustrated in Fig.~\ref{pairs}.  Often
only one of these letters will appear in the subscript; in such a case the letter denotes the
directions and not the chiralities, with the latter being clear from the context.

\subsubsection{Simple 4-point functions}

A 4-point diagram is said to be {\it simple} with respect to a particular pairing of
external lines when it cannot be separated into two parts, each having one pair of the
original external lines, by cutting two internal fermion lines in the 4-point
diagram.  We denote the sum of all 4-point diagrams that are simple with respect to
one specific pairing of external lines by $U$ and refer to it as a {\it simple 4-point
function}. We use $T$ to denote the sum of all 4-point diagrams that are simple with
respect to {\it all but} a specific pairing of external lines. Then the 4-point function, $C$, is the sum
of a $U$ and a $T$ associated with a specific pairing:
\begin{equation}
C = U + T.
\label{v2}
\end{equation}
(For a  while we shall
not distinguish the different chiralities.) There are three ways of pairing off four external lines. One can therefore define
three $U$'s and three $T$'s.  Any particular 4-point diagram is either nonsimple with
respect to only one of these pairings or it is simple with respect to all three
pairings.  It thus follows that $C$ can be written as the sum of three $T$'s (one for
each possible pairing) and $K$, which is the sum of all diagrams that are simple with
respect to all pairings:
\begin{equation}
C = T + T + T + K .
\label{v3}
\end{equation}
The $T$'s are the various bubble chains that can be resummed in closed form.

Once we distinguish between left-handed and right-handed fermions we have the five
$C$'s defined in Sec.~\ref{over} below (\ref{o10}) (see also Fig.~\ref{c's}). The
chirality preserving 4-point functions are $C_0$, $C_R$ and  $C_L$ and the chirality
changing 4-point functions are $C_C$ and $C_C^{\dag}$.

\begin{figure} 
\centerline{\epsfysize = 8 cm \epsfbox{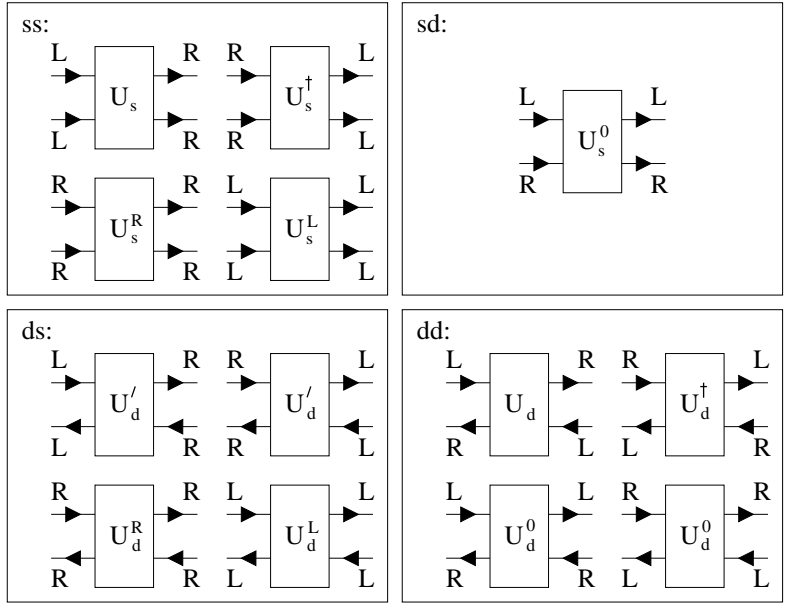}}
\caption{The definition of the simple 4-point functions ($U$'s) in terms of the
pairings and chiralities of the external lines. The four groups indicate the four
different types of fermion pairs.}
\label{u's}
\end{figure}

Each simple 4-point function is defined with respect to one type of fermion pair. The
four categories of $U$'s are shown in Fig.~\ref{u's}. In summary:\
for the {\it ss}-fermion pairs we have $U_s$, $U_s^{\dag}$, $U_s^R$ and $U_s^L$; there
is only one for the {\it sd}-fermion pairs, namely $U_s^0$; those for the {\it
ds}-fermion pairs are $U_d^{\prime}$, $U_d^R$ and $U_d^L$; and the {\it dd}-fermion
pairs are $U_d$, $U_d^{\dag}$ and $U_d^0$. Note that $U_d^{\prime}$ and $U_d^0$ appear
twice each in Fig.~\ref{u's}. There is a $T$ for every $U$. They are related by
(\ref{v2}) for each specific pairing of the external lines of the particular $C$.

It will be useful to define a set of bubble sums\footnote{Note that the first term in
these `bubble sums' is the identity element. One can think of this identity element as
an operator that implies a direct connection of whatever is on either side of it,
when placed inside another diagram.} that are formed as the sum of sequences of the
same $U$ or $C$.  There are only some of these $U$'s and $C$'s that can form such
sequences. The bubble sums involving $C$'s are given in (\ref{ap1}) in
Sec.~\ref{over}. Those involving $U$'s are
\begin{mathletters}
\label{ap2}
\begin{eqnarray}
R_s^U & = & \left[1-\frac{1}{2} U_s^R \right]_{ss}^{-1} \label{ap2rs} \\
L_s^U & = & \left[1-\frac{1}{2} U_s^L \right]_{ss}^{-1} \label{ap2ls} \\
R^U & = & \left[1-U_d^R \right]_{ds}^{-1} \label{ap2r} \\
L^U & = & \left[1-U_d^L \right]_{ds}^{-1} \label{ap2l} \\
B & = & \left[1-U_d^0 \right]_{dd}^{-1} \label{ap2b} .
\end{eqnarray}
Although one can also define such an object for the {\it sd}-fermion
pairs, it does not appear often enough to merit its definition.\end{mathletters} The factor of
$\frac{1}{2}$ that appears in (\ref{ap1rs}), (\ref{ap1ls}), (\ref{ap2rs}) and
(\ref{ap2ls}) removes the overcounting due to the identical fermion lines in the cases
with {\it ss}-fermion pairs. When we encounter this symmetry again it will be referred to as the
{\it ss-symmetry}.

There are other types of transformations which play an important role in some of the
derivations of the subsequent sections, especially in Sec.~\ref{derive}. For this
purpose we define the following transformations, which are denoted by superscripts.
Assume that $G$ is an object with four external lines paired off as $(1,2)$ and
$(3,4)$. Then one can define the following three transformations:
\begin{mathletters}
\label{v4}
\begin{eqnarray}
\left[ G(1,2;3,4) \right]^T & = & G(1,3;2,4) \label{v4a} \\
\left[ G(1,2;3,4) \right]^F & = & G(2,1;4,3) \label{v4b} \\
\left[ G(1,2;3,4) \right]^R & = & G(3,4;1,2) \label{v4c} .
\end{eqnarray}
The last one is referred to as a back-to-front transformation. 
\end{mathletters} 

\subsection{The T-terms}
\label{nonint}

\subsubsection{The master equations}

Each master equation
expresses a 4-point function ($C$) in terms of some simple 4-point function ($U$) and
itself, and leads to an expression for the $U$ in terms of the $C$. The $T$-term then
follows as the difference between the $C$ and this $U$. We have four types of fermion pairs,
three of which have four master equations each.  The remaining fermion pair type has only one
master equation, and we start our discussion with this equation.

$C_0$ is the only $C$ that can have the {\it sd}-fermion pair type.
The associated
simple 4-point function is $U_s^0$ and the master equation is (see Fig.~\ref{master}):
\begin{equation}
C_0 = U_s^0 + U_s^0 C_0 .
\label{ap3}
\end{equation}
From (\ref{ap3}) it then directly follows that 
\begin{equation}
U_s^0 = C_0 [1 + C_0]_{sd}^{-1} .
\label{ap4}
\end{equation}
The $T$-term in the gap equation for $C_0$ is now given 
by
\begin{equation}
T_s^0 = C_0 - U_s^0 = C_0^2 [1 + C_0]_{sd}^{-1} .
\label{ap5}
\end{equation}

For the other three types of fermion pairs we first present all the master equations. 
The master equations for the {\it ss}-fermion pairs are:
\begin{mathletters}
\label{ap6}
\begin{eqnarray}
C_C & = & U_s + \frac{1}{2} U_s^L C_C + \frac{1}{2} U_s C_R \label{ap6c} \\
C_C^{\dag} & = & U_s^{\dag} + \frac{1}{2} U_s^R C_C^{\dag} + \frac{1}{2} U_s^{\dag}
C_L \label{ap6cc} \\
C_R & = & U_s^R + \frac{1}{2} U_s^R C_R + \frac{1}{2} U_s^{\dag} C_C \label{ap6r} \\
C_L & = & U_s^L + \frac{1}{2} U_s^L C_L + \frac{1}{2} U_s C_C^{\dag} \label{ap6l} .
\end{eqnarray}
The factor, $\frac{1}{2}$, is due to the {\it ss}-symmetry mentioned at the end of
Sec.~\ref{algebra}. \end{mathletters} Next we have the master equations for the {\it
ds}-fermion pairs:
\begin{mathletters}
\label{ap7}
\begin{eqnarray}
C_0 & = & U_d^{\prime} + U_d^R C_0 + U_d^{\prime} C_L \label{ap7a} \\
C_0 & = & U_d^{\prime} + U_d^L C_0 + U_d^{\prime} C_R \label{ap7b} \\
C_R & = & U_d^R + U_d^R C_R + U_d^{\prime} C_0 \label{ap7r} \\
C_L & = & U_d^L + U_d^L C_L + U_d^{\prime} C_0 \label{ap7l} .
\end{eqnarray}
The first two equation, (\ref{ap7a}) and (\ref{ap7b}), and the last terms in the last
two equation, (\ref{ap7r}) and (\ref{ap7l}), may seem ambiguous. \end{mathletters}
Note, however, that corresponding external lines of different terms in an equation must
have the same directions and chiralities. As a result one can see that (\ref{ap7a}) and
(\ref{ap7b}) are related by a back-to-front transformation, as defined in (\ref{v4c})
and that the fermion lines which connect $C_0$ and $U_d^{\prime}$ are left-handed in
(\ref{ap7r}) and right-handed in (\ref{ap7l}). 

The final set of master equations are for the {\it dd}-fermion pairs:
\begin{mathletters}
\label{ap8}
\begin{eqnarray}
C_C & = & U_d + U_d C_0 +U_s^0 C_C \label{ap8c} \\
C_C^{\dag} & = & U_d^{\dag} + U_d^{\dag} C_0 + U_d^0 C_C^{\dag} \label{ap8cc} \\
C_0 & = & U_d^0 + U_d^0 C_0 + U_d^{\dag} C_C \label{ap8a} \\
C_0 & = & U_d^0 + U_d^0 C_0 + U_d C_C^{\dag} \label{ap8b} .
\end{eqnarray}
The last term in each of (\ref{ap8a}) and (\ref{ap8b}) define  the chirality
assignments on the external lines and thence the definition of the other terms follow
unambiguously.
\end{mathletters}

\subsubsection{Simple 4-point functions}

The master equations are manipulated to give the $U$'s expressed in terms of only the
$C$'s. The
expressions of the $U$'s are required for the pair cycle terms of the effective action which
are derived in Sec.~\ref{paircycles}. From the twelve expressions in (\ref{ap6}), (\ref{ap7}) and (\ref{ap8}) one can write
down simpler expressions using the definitions in (\ref{ap1}) and  (\ref{ap2}). First
we consider the {\it ss}-fermion pairs in detail. With the use of the definitions in
(\ref{ap1rs}), (\ref{ap1ls}),  (\ref{ap2rs}) and (\ref{ap2ls}) the expressions in
(\ref{ap6c}) and (\ref{ap6cc}) become
\begin{mathletters}
\label{ap9}
\begin{equation}
C_C R_s = L_s^U U_s
\label{ap9rs}
\end{equation}
and
\begin{equation}
C_C^{\dag} L_s = R_s^U U_s^{\dag} .
\label{ap9ls}
\end{equation}
Using the same definitions one can express the next two equations,
(\ref{ap6r}) and (\ref{ap6l}) as
\begin{equation}
1 = R_s^U R_s + \frac{1}{4} R_s^U U_s^{\dag} C_C R_s = R_s^U R_s + \frac{1}{4}
C_C^{\dag} L_s C_C R_s
\label{ap9ru}
\end{equation}
and
\begin{equation}
1 = L_s^U L_s + \frac{1}{4} L_s^U U_s C_C^{\dag} L_s = L_s^U L_s + \frac{1}{4}
C_C R_s C_C^{\dag} L_s ,
\label{ap9lu}
\end{equation}
where we have used the identities in (\ref{ap9rs}) and (\ref{ap9ls}).
\end{mathletters} From  (\ref{ap9ru}) and (\ref{ap9lu}) one can now write down the
expressions  for $U_s^R$ and $U_s^L$:
\begin{mathletters}
\label{ap10a}
\begin{equation}
U_s^R = 2 - 2 R_s \left[ 1 - \frac{1}{4} C_C^{\dag} L_s C_C R_s \right]_s^{-1} 
\label{ap10}
\end{equation}
and
\begin{equation}
U_s^L = 2 - 2 L_s \left[ 1 - \frac{1}{4} C_C R_s C_C^{\dag} L_s \right]_s^{-1}
\label{ap11} .
\end{equation}
One can eliminate $R_s^U$ between (\ref{ap9ls}) and (\ref{ap9ru}) to find an 
expression for $U_s^{\dag}$,
\begin{equation}
U_s^{\dag} = R_s C_C^{\dag} L_s \left[1 - \frac{1}{4} C_C R_s C_C^{\dag} L_s
\right]_s^{-1} 
\label{ap12}
\end{equation}
and similarly one can eliminate $L_s^U$ between (\ref{ap9rs}) and (\ref{ap9lu}) 
to find an expression for $U_s$,
\begin{equation}
U_s = L_s C_C R_s \left[1 - \frac{1}{4} C_C^{\dag} L_s C_C R_s \right]_s^{-1} .
\label{ap13}
\end{equation}
In the derivation of these equations we have used the identity \end{mathletters}
\begin{equation}
A \left[ 1 - B A \right]^{-1} = \left[ 1 - A B \right]^{-1} A .
\label{ap14}
\end{equation}

The {\it ds}-fermion pairs and {\it dd}-fermion pairs follow the same steps. The
equivalent expressions for the {\it ds}-fermion pairs are:
\begin{mathletters}
\label{ap15}
\begin{eqnarray}
C_0 L & = & R^U U_d^{\prime} \label{ap15l} \\
C_0 R & = & L^U U_d^{\prime} \label{ap15r}
\end{eqnarray}
\begin{eqnarray}
1 & = & R^U R + R^U U_d^{\prime} C_0 R = R^U R + C_0 L C_0 R
\label{ap15ru} \\ 1 & = & L^U L + L^U U_d^{\prime} C_0 L = L^U L + C_0 R C_0 L .
\label{ap15lu}
\end{eqnarray}
From these we derive the expressions for the following $U$'s: \end{mathletters}
\begin{mathletters}
\label{ap16a}
\begin{eqnarray}
U_d^R & = & 1 - R \left[ 1 - C_0 L C_0 R \right]_d^{-1} \label{ap16} \\
U_d^L & = & 1 - L \left[ 1 - C_0 R C_0 L \right]_d^{-1} \label{ap17} \\
U_d^{\prime} & = &  R C_0 L\left[1 - C_0 R C_0 L \right]_d^{-1} \label{ap18} \\
U_d^{\prime} & = & L C_0 R \left[1 - C_0 L C_0 R \right]_d^{-1} \label{ap19} .
\end{eqnarray}
The last two equations, (\ref{ap18}) and (\ref{ap19}), are related through a 
back-to-front transformation. \end{mathletters} The expressions for the {\it
dd}-fermion pairs are:
\begin{mathletters}
\label{ap20}
\begin{eqnarray}
C_C Z & = & B U_d \label{ap20c} \\
C_C^{\dag} Z & = & B U_d^{\dag} \label{ap20cc}
\end{eqnarray}
\begin{eqnarray}
1 & = & B Z + B U_d^{\dag} C_C Z = B Z + C_C^{\dag} Z C_C Z \label{ap20z} \\
1 & = & B Z + B U_d C_C^{\dag} Z = B Z + C_C Z C_C^{\dag} Z \label{ap20zc} ,
\end{eqnarray}
and from them follow the expressions for the remaining $U$'s: \end{mathletters}
\begin{mathletters}
\label{ap21a}
\begin{eqnarray}
U_d^0 & = & 1 - Z \left[ 1 - C_C Z C_C^{\dag} Z \right]_d^{-1} \label{ap21} \\
U_d^0 & = & 1 - Z \left[ 1 - C_C^{\dag} Z C_C Z \right]_d^{-1} \label{ap22} \\
U_d & = & Z C_C Z \left[1 - C_C^{\dag} Z C_C Z \right]_d^{-1} \label{ap23} \\
U_d^{\dag} & = & Z C_C^{\dag} Z \left[1 - C_C Z C_C^{\dag} Z \right]_d^{-1} .
\label{ap24}
\end{eqnarray}
Here the first two equations, (\ref{ap21}) and (\ref{ap22}), are related through 
a back-to-front transformation. \end{mathletters}

\subsubsection{T-term expressions}

The final step in the derivation of the $T$-terms in the gap equation is to use equations
of the form $T=C-U$ to find their expressions. The expressions for the $U$'s are those
provided in (\ref{ap4}), (\ref{ap10a}), (\ref{ap16a}) and (\ref{ap21a}). This step is
quite straight forward so we merely quote all the relevant expressions.

{\it ss}-fermion pairs:

\begin{mathletters}
\label{ap25}
\begin{eqnarray}
T_s & = & C_C - U_s = \frac{1}{2} U_s^L C_C + \frac{1}{2} U_s C_R = C_C - L_s C_C R_s
\left[ 1 - \frac{1}{4} C_C^{\dag} L_s C_C R_s \right]_s^{-1} \label{ap25a} \\
T_s^{\dag} & = & C_C^{\dag} - U_s^{\dag} = \frac{1}{2} U_s^R C_C^{\dag} + \frac{1}{2}
U_s^{\dag} C_L = C_C^{\dag} - R_s C_C^{\dag} L_s \left[ 1 - \frac{1}{4} C_C R_s
C_C^{\dag} L_s \right]_s^{-1} \label{ap25b} \\
T_s^R & = & C_R - U_s^R = \frac{1}{2} U_s^R C_R + \frac{1}{2} U_s^{\dag} C_C = C_R + 2
R_s \left[ 1 - \frac{1}{4} C_C^{\dag} L_s C_C R_s \right]_s^{-1} - 2 \label{ap25c} \\
T_s^L & = & C_L - U_s^L = \frac{1}{2} U_s^L C_L + \frac{1}{2} U_s C_C^{\dag} = C_L + 2
L_s \left[ 1 - \frac{1}{4} C_C R_s C_C^{\dag} L_s \right]_s^{-1} - 2 \label{ap25d}
\end{eqnarray}
\end{mathletters}

{\it ds}-fermion pairs:

\begin{mathletters}
\label{ap26}
\begin{eqnarray}
T_d^{\prime} & = & C_0 - U_d^{\prime} = U_d^R C_0 + U_d^{\prime} C_L = C_0 - R C_0 L
\left[ 1 - C_0 R C_0 L \right]_d^{-1} \label{ap26a} \\
T_d^{\prime} & = & C_0 - U_d^{\prime} = U_d^L C_0 + U_d^{\prime} C_R = C_0 - L C_0 R
\left[ 1 - C_0 L C_0 R \right]_d^{-1} \label{ap26b} \\
T_d^R & = & C_R - U_d^R = U_d^R C_R + U_d^{\prime} C_0 = C_R + R \left[ 1 - C_0 L
C_0 R \right]_d^{-1} - 1 \label{ap26c} \\
T_d^L & = & C_L - U_d^L = U_d^L C_L + U_d^{\prime} C_0 = C_L + L \left[ 1 - C_0 R
C_0 L \right]_d^{-1} - 1 \label{ap26d}
\end{eqnarray}
\end{mathletters}

{\it dd}-fermion pairs:

\begin{mathletters}
\label{ap27}
\begin{eqnarray}
T_d  & = & C_C - U_d = U_d C_0 +U_s^0 C_C = C_C - Z C_C Z \left[ 1 - C_C^{\dag}
Z C_C Z \right]_d^{-1} \label{ap27a} \\
T_d^{\dag}  & = & C_C^{\dag} - U_d^{\dag} = U_d^{\dag} C_0 + U_d^0 C_C^{\dag} =
C_C^{\dag} - Z C_C^{\dag} Z \left[ 1 - C_C Z C_C^{\dag} Z \right]_d^{-1} \label{ap27b}
\\ T_d^0  & = & C_0 - U_d^0 = U_d^0 C_0 + U_d^{\dag} C_C = C_0 + Z \left[ 1 -
C_C^{\dag} Z C_C Z \right]_d^{-1} - 1 \label{ap27c} \\ 
T_d^0  & = & C_0 - U_d^0 = U_d^0 C_0 + U_d C_C^{\dag} = C_0 + Z \left[ 1 - C_C Z
C_C^{\dag} Z \right]_d^{-1} - 1 \label{ap27d}
\end{eqnarray}
\end{mathletters}

A back-to-front transformation relates (\ref{ap26a}) and (\ref{ap26b}), as well as
(\ref{ap27c}) and (\ref{ap27d}).

\subsection{4-particle irreducible diagrams}
\label{4part}

The $K$-term contains all 4-point diagrams that are simple with respect to all
pairings of external lines. We want to express this sum of diagrams in terms of
the $C$'s. That means that we have to recast the set of diagrams in
the $K$-term such that all the 4-point subdiagrams inside these diagrams are contained
inside $C$'s. This process is neatly performed by constructing all the
4-particle irreducible (4PI) diagrams, using the $C$'s as 4-fermion vertices, together
with the full fermion propagator and the other vertices and propagators of a gauge
theory. First we give the formal definition of a 4PI diagram. Then we clarify this
definition.

\begin{figure}
\centerline{\epsfysize = 5 cm \epsfbox{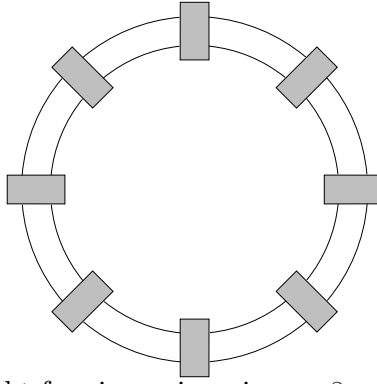}}
\caption{A pair cycle with eight fermion pairs -- i.e.\ an 8-cycle. The shaded
rectangles denote simple subdiagrams.}
\label{paircycle}
\end{figure}

4PI 4-point diagrams are defined by first defining 4PI vacuum diagrams. The 4PI 4-point
diagrams are then obtained from these vacuum diagrams by removing one 4-fermion vertex
($C$). The definition of a 4PI vacuum diagram will make use of the concept of a {\it pair
cycle}, where {\it pair} refers to a pair of fermion lines.  A {\it pair cycle} is a
loop of pairs of fermion lines that connect simple subdiagrams to form a vacuum
diagram. An example is shown in Fig.~\ref{paircycle}.  A pair cycle with $\mu$ pairs
is referred to as a {\it $\mu$-cycle}. Cutting the lines of all $\mu$ pairs in a
$\mu$-cycle would separate the vacuum diagram into $\mu$ disconnected simple 4-point
functions.  If one and only one of the simple subdiagrams in a $\mu$-cycle is just a
4-fermion vertex ($C$), the $\mu$-cycle is said to be a {\it trivial} $\mu$-cycle.

\begin{figure}
\centerline{\epsfysize = 5 cm \epsfbox{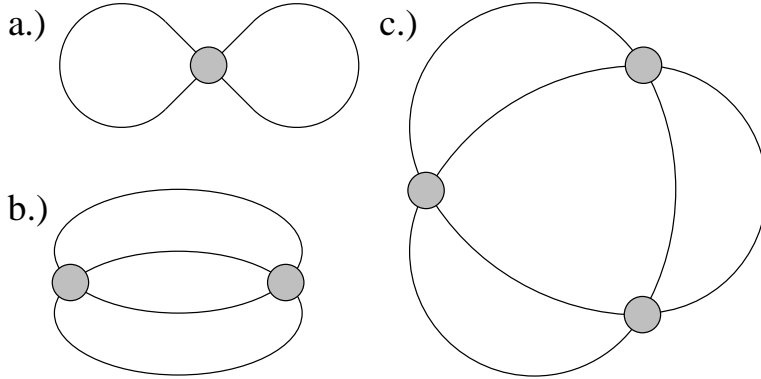}}
\caption{Three vacuum diagrams that are not 4PI. The shaded circles
denote 4-point vertices ($J$'s or $C$'s).}
\label{4pi}
\end{figure}

The formal definition of a 4PI vacuum diagram is then that {\it all the pair cycles in
a 4PI vacuum diagram are either nontrivial 1-cycles or trivial 2-cycles}.\footnote{We
deviate slightly from the definition in \cite{r_dm} in that their definition include
trivial 1-cycles as 4PI diagrams. The only diagram with a trivial one cycle is shown
in Fig.~\ref{4pi}a. This diagram has the undesirable feature that, upon removing the
4-fermion vertex, it gives the disconnected propagators which are excluded from $C$'s
definition as shown in (\ref{ei9}).}

Let's clarify this definition by considering the requirements for 4PI diagrams and
giving motivations for these requirements. For 2PI vacuum diagrams one simply requires
that it should not be possible to separate the diagram into two parts by cutting two
different fermion lines in the diagram. Generalizing this rule to four fermion lines
for 4PI diagrams is not enough, because then no 4PI diagram would contain a $C$. So
one must add that if by cutting four different fermion lines in a diagram, it
separates into two parts with one and only one part being a 4-fermion vertex ($C$), then
the diagram is 4PI. Thus we allow trivial two-cycles.
A vacuum diagram that consists of two $C$'s with their lines connected to
each other, as shown in Fig.~\ref{4pi}b, produces the 4-fermion vertex ($C$) upon removal
of one of the $C$'s. The latter 4-point diagram is by definition not part of $K$, and
thus nontrivial 2-cycles are excluded in the definition of 4PI vacuum diagrams.

We could have tried to define 4PI vacuum diagrams as all vacuum diagrams that cannot
be separated into two disconnected parts by cutting four fermion lines, unless one and
only one of the parts is just the 4-fermion vertex, $C$. However, this definition
still includes the diagram consisting of three $C$'s, shown in Fig.~\ref{4pi}c. This
diagram must be excluded because it becomes a diagram in $T$ when one removes one of
the $C$'s. Thus we see why the definition of 4PI vacuum diagrams must exclude all
diagrams with 3-cycles and higher. 

Now we can obtain all the 4PI 4-point diagrams from the 4PI vacuum diagrams by cutting
out (or opening up) one 4-fermion vertex ($C$). As an example:\ if a 4PI vacuum
diagram contains five 4-fermion vertices then one can form five 4PI 4-point diagrams
from it. If two or more of these 4PI 4-point diagrams are identical the original
vacuum diagram would have a symmetry factor which cancels the overcounting.

The amputated 4PI 4-point diagrams that are constructed from the $C$'s, the full
fermion propagators and the other vertices and propagators of a gauge theory, are
exactly those which are contained in the $K$-term.  To see this, remember that the
$C$'s represent the sets of all connected amputated 4-point diagrams with the
appropriate chiralities on their external lines. By replacing the $C$'s inside the 4PI
diagrams with these sets of 4-point diagrams, one generates all the 4-point diagrams
which are simple with respect all pairings of external lines.\footnote{Note that,
although a 4PI 4-point diagram implies a diagram that is simple with respect to all
pairings of external lines, the converse is not necessarily true.} Amputating these
simple diagrams, one obtains the set of diagrams which form $K$.
\begin{equation}
K[C] = \left[{\delta V_{4PI}[C] \over \delta C}\right]_{amp}
\label{v5}
\end{equation}
$V_{4PI}[C]$ denotes the sum of all 4PI vacuum diagrams containing the 4-point
functions ($C$) as 4-point vertices.

\subsection{The gap equations}
\label{express}

The expansions for the five 4-point functions in terms of the $T$'s and $K$'s are:
\begin{mathletters}
\label{v6}
\begin{eqnarray}
C_C & = & T_s + T_d + (T_d)^T + K_C \label{v6a} \\
C_C^{\dag} & = & T^{\dag}_s + T^{\dag}_d + (T^{\dag}_d)^T + K_C^{\dag} \label{v6b}
\\ C_0 & = & T^0_s + T^0_d + T^{\prime}_d + K_0 \label{v6c} \\
C_R & = & T^R_s + T^R_d + (T^R_d)^T + K_R \label{v6d} \\
C_L & = & T^L_s + T^L_d + (T^L_d)^T + K_L . \label{v6e}
\end{eqnarray}
Note that there are always two of the $T$'s with subscript $d$ and only one $T$ with
subscript $s$. \end{mathletters} The superscript $T$ indicates that the two incoming or
two outgoing lines are interchanged. There is only one of the equations where the two
$T$'s with subscript $d$ are different -- the equation for $C_0$.  The distinction is
that for $T^0_d$ the chiralities in a pair are opposite and for $T^{\prime}_d$ the
chiralities in a pair are the same. (This is the same as for $U^0_d$ and
$U^{\prime}_d$, shown in Fig.~\ref{u's}.)

When distinguishing the chiralities, one has five different 4-point functions on which
$V_{4PI}$ depends. The $K$-term in each specific expression in (\ref{v6}) is generated
by taking the functional derivative of $V_{4PI}$ with respect to the appropriate
4-point function. (For $K$ ($K^{\dag}$) one must take the functional
derivative with respect to $C^{\dag}$ ($C$).) The expressions for all the $T$'s are provided
in (\ref{ap5}), (\ref{ap25}), (\ref{ap26}) and (\ref{ap27}). These expressions, together with
the definitions in (\ref{ap1}), now complete the gap equations in (\ref{v6}). 

Next we derive the effective action, which leads to the same gap equations.

\section{Derivation of the effective action}
\label{effek}

We must find an expression for the effective action in terms of the 4-point
functions ($C$'s) as outlined
in Sec.~\ref{green}. It requires the use of the 4-fermion sources, $J$'s, which are
shown in (\ref{ei6}).\footnote{In  \cite{r_dm} nonlocal
vertices, denoted by $v$'s, are used to represent all interactions in the
theory. We replace $v$'s by sources,
$J$'s, and keep them separate from the gauge interactions in our analysis.} We consider all the
connected 2PI vacuum diagrams that can be generated with these sources treated as 4-fermion
vertices, our full fermion propagator, and other vertices and propagators of a gauge theory. The
sum of all these vacuum diagrams is the generating functional, $W(S,J)$. The
effective action is obtained by performing a Legendre
transform, which replaces the source dependence by a dependence on the 4-point functions
($C$'s). The challenge is therefore to express $W$ in terms of $C$'s, which only implicitly
depend on the $J$'s.

In the sections below we show how this is achieved with the aid of a topological
equation, which allows us to avoid overcounting diagrams. The various terms in the
topological equation are explained below.

\subsection{Topological equation}
\label{sec:topol}

The topological equation,
\begin{equation}
1 = N_{skel} - N_{art} + N_{vert} + \sum_{pair~cycles} [1 - N_p + N_{pp}] ,
\label{e1}
\end{equation}
which is proved in \cite{r_dm}, is an index theorem for different numbers of elements
which are present in vacuum diagrams. The quantities on the right-hand side, which we
shall define shortly, indicate the number of particular elements which are present in
any specific vacuum diagram.  Thus for any vacuum diagram, if one determines the
number of each type of element which it contains, then adds or subtracts these
numbers in the way indicated in (\ref{e1}), the result is equal to unity. (See also
the discussion in Sec.~\ref{over}.)

The idea now is to find the various ways the various elements can be identified in each
vacuum diagram, and then collect together the sets of identifications for each type of
element. Carrying this out for all vacuum diagrams gives
\begin{equation}
W = W_{skel} - W_{art} + W_{vert} + W_c - W_p + W_{pp} .
\label{e2}
\end{equation}
Every diagram on the left-side of (\ref{e2}) appears various numbers of times in each of the
various terms on the right-hand side, as specified in (\ref{e1}). The point is that each of the
terms on the right-hand side has an unambiguous expression in terms of the 4-point functions.

This general procedure is now applied to the case where one distinguishes the
chiralities on the fermion lines. Care must be taken to ensure that any overcounting
that may result from the symmetries associated with the chiralities is removed. Except
for the $C_0$, all 4-point functions have the same chiralities on both incoming, as
well as on both outgoing fermion lines.  There is then an associated symmetry factor of
$\frac{1}{4}$, as revealed in the
expression in (\ref{ei6}).

\begin{figure}
\centerline{\epsfysize = 3 cm \epsfbox{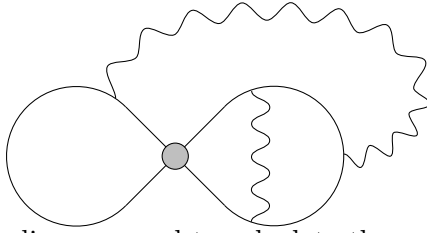}}
\caption{A non-4PI vacuum diagram used to calculate the number of particular elements
of a vacuum diagram for the topological equation.}
\label{fig:topol}
\end{figure}

We first consider the first three terms in the topological equation, while last three terms are
discussed in Sec.~\ref{paircycles}. We use as an example, the vacuum diagram shown in
Figure~\ref{fig:topol}. We shall determine, for each term in the topological equation, the
number of times that particular element appears in this diagram.

A 4PI {\it skeleton} is what remains of a 4PI vacuum diagram after one has cut out all
the 4-fermion vertices.  A vacuum diagram that is not 4PI does in general contain 4PI
skeletons by cutting out appropriate subdiagrams, and $N_{skel}$ counts the number of such
skeletons. The two gauge exchanges in our example are the only 4PI skeletons, because in
each case, if one replaces the rest of the diagram with one 4-fermion vertex, the result is
a 4PI diagram. By doing this replacement on any other 4-point subdiagram one finds
that there are no other 4PI skeletons. So for our example we have $N_{skel}=2$.

The term, $W_{skel}$, is generated by the sum of all 4PI vacuum diagrams, $V_{4PI}$,
where all possible connected 4-point diagrams (the $C$'s) replace the 4-fermion
vertices. This gives
\begin{equation}
W_{skel} = V_{4PI}[C,C^{\dag},C_0,C_R,C_L] ,
\label{e3}
\end{equation}
The $C$'s now play the role
of 4-fermion vertices, in place of the $J$'s.

The number of articulated quartets, $N_{art}$, counts the number of sets of four
fermion lines in a vacuum diagram that one can cut to separate the vacuum diagram
into two disconnected parts. There are three such sets in our example. Two sets
connect the two gauge exchanges to the rest of the diagram and one set connects the
4-fermion vertex to the rest of the diagram. Thus we have $N_{art}=3$ for our example.

One can generate the term, $W_{art}$, for the set of all vacuum diagrams by closing
off the lines of one 4-point diagram by those of another 4-point diagram for all
possible connected 4-point diagrams.  The result is just the trace over the product of
two $C$'s,
\begin{equation}
W_{art} = {\rm Tr} \left \{ \frac{1}{4} C C^{\dag} + \frac{1}{8} C_R^2 + \frac{1}{8}
C_L^2 + \frac{1}{2} C_0^2 \right \} .
\label{e4}
\end{equation}
The symmetry factors appear as follows:\ if the term is
made of 4-point functions that have identical incoming or outgoing lines  then the same
factor of $\frac{1}{4}$ which appears in (\ref{ei6}) accompanies  this term -- this is
the case for the first three terms in (\ref{e4}); if the two 4-point functions in a
term are the same (as in the last three terms of (\ref{e4})) we need an extra factor
of $\frac{1}{2}$.

The third term in (\ref{e1}), $N_{vert}$, is just the number of 4-fermion vertices
in a vacuum diagram. Our example contains only one such vertex, so
$N_{vert}=1$.

The term, $W_{vert}$, is generated by closing the four lines of all unamputated
4-point diagrams (including disconnected ones) off with the appropriate source $J$, which
gives a trace over the product of a $J$ with the Green function ($G$). This
generates the five source terms which are removed by the Legendre transform, (\ref{ei7}):
\begin{equation}
W_{vert} = \frac{1}{4} J G^{\dag} + \frac{1}{4} J^{\dag} G + J_0 G_0 + \frac{1}{4} J_R
G_R + \frac{1}{4} J_L G_L .
\label{e5}
\end{equation}
Here we again have factors of $\frac{1}{4}$ which appear due to identical incoming and
outgoing lines.

\subsection{Pair cycles}
\label{paircycles}

A pair cycle is a loop of pairs of fermion
lines that connect simple subdiagrams to form a vacuum diagram. (See
Fig.~\ref{paircycle}.) The sum over pair cycles makes up the last three terms in the
topological equation (\ref{e1}). The first term in the sum, $\sum 1$, just counts the
number of such pair cycles in a vacuum diagram. The next term, $\sum N_p$, counts the
number of pairs in each pair cycle of the vacuum diagram and the third term in the
sum, $\sum N_{pp}$, counts the pairs of pairs in each pair cycle of the diagram. One
can see that for 1-cycles and 2-cycles these three terms add up to zero. Therefore it
is only for pair cycles with three or more pairs that these terms make a nontrivial
contribution to the topological equation. In our example there is only one pair cycle
with more than two pairs and it is a 3-cycle. This gives $\sum(1-N_p+N_{pp}) = 1-3+3 =
1$ for our example. Adding all the numbers for our example according to (\ref{e1})
then gives 1 as it should.

One can define the fourth term in (\ref{e2}), $W_c$, as the trace of a power series of
simple 4-point functions ($U$'s) with the appropriate symmetry factors.  A product of
$n$ simple 4-point functions forms a chain of $n$ fermion pairs and the trace of this
product closes off the chain into a pair cycle. The next term $W_p$ can be generated
from the previous one $W_c$ by including a parameter $g$ with each $U$ and then taking
the derivative with respect to $g$. This counts the number of pairs in each pair cycle, and
the second derivative gives twice the number of pairs of pairs. Half of
the latter gives the last term, $W_{pp}$. Since the $U$'s can be expressed in terms of the
4-point functions ($C$'s), this procedure leads to the required expressions for the three pair
cycle terms. The pair cycles found in our analysis are more complicated than those in
\cite{r_dm} because there are several ways to form them due to the distinguished chiralities.

We follow the approach of Sec.~\ref{algebra}, which is to divide the different
$U$'s into four groups depending on their type of fermion pairs. The pair cycle terms
are expressed as
\begin{equation}
H_{Total} = W_c - W_p + W_{pp}
\label{b0}
\end{equation}
for each type of fermion pairs. The aim is to find all the different pair cycles by
identifying  all the $U$'s that can form such cycles. Some $U$'s can form pair cycles
on their own.  The two pairs of external lines on these $U$'s are compatible, which
implies that two of these $U$'s can be interconnected. The other $U$'s have
incompatible pairs. Although they cannot form pair cycles on their own, they can be 
combined with other $U$'s to form a group of simple 4-point functions which can be
used to form  pair cycles.  Such groups of simple 4-point functions are denoted by
$M$'s. We shall discuss  each of the type of fermion pairs in turn, starting with the
simplest case.

\subsubsection{{\it sd}-fermion pairs}

The {\it sd}-fermion pair has only one simple 4-point function:\ $U_s^0$.  There are no
simple 4-point function groups, $M$'s, for this type of fermion pair. The pair cycle
is formed by defining the following quantity:
\begin{equation} 
A_{sd}(g) = {\rm Tr} \left \{ \sum_{n=1}^{\infty} {1\over n} (g U_s^0)^n \right 
\} = - {\rm Tr} \left \{ \ln \left( 1 - g U_s^0 \right)_s \right \} .
\label{b1} 
\end{equation} 
Setting $g=1$ in (\ref{b1}) one finds the sum of all pair cycles with this type of
fermion pair.  The number of pairs per cycle is obtained by taking the derivative of
$A_s^0(g)$ with respect to $g$, while a second derivative gives twice the number of
pairs of pairs. The sum of pair cycle terms, (\ref{b0}), with {\it sd}-fermion pairs are
then given by
\begin{eqnarray} 
H_{sd} & = & \left[ A_{sd}(g) - {d \over d g} A_{sd}(g) + \frac{1}{2} {d^2 \over d
g^2} A_{sd}(g) \right]_{g=1} \nonumber \\ & = & {\rm Tr} \left \{ - \ln \left( 1
- U_s^0 \right)_s - \left[ 1 - U_s^0 \right]_s^{-1} U_s^0 + \frac{1}{2} \left[ 1 - 
U_s^0 \right]_s^{-1} U_s^0 \left[ 1 - U_s^0 \right]_s^{-1} U_s^0 \right \} .
\label{b2} 
\end{eqnarray}
From (\ref{ap3}) one can show that 
\begin{equation}
\left[ 1 - U_s^0 \right]_s^{-1} U_s^0 = C_0 .
\label{b3}
\end{equation}
Using (\ref{b3}) and (\ref{ap4}), (\ref{b2}) becomes
\begin{equation} 
H_{sd} = {\rm Tr} \left \{ \ln \left(1 + C_0 \right)_s - C_0 + \frac{1}{2} C_0^2
\right \} .
\label{b4}
\end{equation}
This is similar to the result in \cite{r_dm}. The subsequent cases 
are all treated according to the same steps. For each pair cycle one can define 
\begin{equation} 
H_f = \left[ A_f(g) - {d \over d g} A_f(g) + \frac{1}{2} {d^2 \over d
g^2} A_f(g) \right]_{g=1} ,
\label{b5} 
\end{equation}
where the subscript $f$ denotes the specific type of simple 4-point function or simple 
4-point function group for that pair cycle. Some algebra
necessary to get this expression into a form that only consist of $C$'s.

\subsubsection{{\it ss}-fermion pairs}

There are two simple 4-point functions that can directly form pair cycles in the {\it
ss}-fermion pairs:\ $U_s^R$ and $U_s^L$. There is also a simple 4-point function group
that can form pair cycles:
\begin{equation} 
M_s(g) = \frac{g^2}{2} \left[1-\frac{g}{2} U^R_s \right]_s^{-1} U_s^{\dag}
\left[1-\frac{g}{2} U^L_s\right]_s^{-1} U_s = \frac{1}{2}g^2 R_s^U(g) U_s^{\dag}
L_s^U(g) U_s .
\label{b6}
\end{equation}

Here the pair cycles are formed by
\begin{equation} 
A_{ss}(g) = {\rm Tr} \left \{ \sum_{n=1}^{\infty} {1\over n} \left( \frac{1}{2}
M(g) \right)^n \right \} = -{\rm Tr} \left \{ \ln \left( 1 - \frac{1}{2} M(g)
\right)_s \right \} ,
\label{b7} 
\end{equation} 
where $M(g)$ denotes $M_s(g)$, $g U_s^R$ or $g U_s^L$. The factor of $\frac{1}{2}$ is
necessary to remove an overcounting which occurs as a result of the {\it ss}-symmetry
as discussed in Sec.~\ref{algebra}.  For $U_s^R$ and $U_s^L$ the expressions are only
slightly more complicated than in the {\it sd}-case. We define separate quantities:
\begin{eqnarray}
H_s^R & = & {\rm Tr} \left \{ - \ln \left( 1 - \frac{1}{2} U_s^R \right)_s -
\frac{1}{2} \left[ 1 - \frac{1}{2} U_s^R \right]_s^{-1} U_s^R  \right.  \nonumber
\\ & & + \left. \frac{1}{8} \left[ 1 - \frac{1}{2} U_s^R \right]_s^{-1} U_s^R 
\left[ 1 - \frac{1}{2} U_s^R \right]_s^{-1} U_s^R \right \}
\label{b8}
\end{eqnarray}
and
\begin{eqnarray}
H_s^L & = & {\rm Tr} \left \{ - \ln \left( 1 - \frac{1}{2} U_s^L \right)_s -
\frac{1}{2} \left[ 1 - \frac{1}{2} U_s^L \right]_s^{-1} U_s^L \right.  \nonumber
\\  & & + \left.  \frac{1}{8} \left[ 1 - \frac{1}{2} U_s^L \right]_s^{-1} U_s^L 
\left[ 1 - \frac{1}{2} U_s^L \right]_s^{-1} U_s^L \right \}
\label{b9}
\end{eqnarray}
From (\ref{ap9ru}) and (\ref{ap9lu}) it can be shown that
\begin{equation}
\left[ 1 - \frac{1}{2} U_s^R \right]_s^{-1} U_s^R = C_R - \frac{1}{2} C_C^{\dag} 
L_s C_C
\label{b10}
\end{equation}
and
\begin{equation}
\left[ 1 - \frac{1}{2} U_s^L \right]_s^{-1} U_s^L = C_L - \frac{1}{2} C_C 
R_s C_C^{\dag} .
\label{b11}
\end{equation}
Using (\ref{b10}) and (\ref{ap10}) one finds that (\ref{b8}) becomes
\begin{eqnarray}
H_s^R & = & {\rm Tr} \left \{ \ln \left( 1 - \frac{1}{4} C_C^{\dag} L_s C_C R_s
\right)_s + \ln \left( 1 + \frac{1}{2} C_R \right)_s - \frac{1}{2} \left( C_R -
\frac{1}{2} C_C^{\dag} L_s C_C \right) \right.  \nonumber \\ & & + \left.
\frac{1}{8} \left(C_R - \frac{1}{2} C_C^{\dag} L_s C_C \right)^2 \right \} .
\label{b12}
\end{eqnarray}
Similarly, from (\ref{b11}) and (\ref{ap11}), (\ref{b9}) becomes
\begin{eqnarray}
H_s^L & = & {\rm Tr} \left \{ \ln \left( 1 - \frac{1}{4} C_C R_s C_C^{\dag} L_s
\right)_s + \ln \left( 1 + \frac{1}{2} C_L \right)_s - \frac{1}{2} \left( C_L -
\frac{1}{2} C_C R_s C_C^{\dag} \right) \right.  \nonumber \\ & & + \left.
\frac{1}{8} \left(C_L - \frac{1}{2} C_C R_s C_C^{\dag} \right)^2 \right \} .
\label{b13}
\end{eqnarray}
Due to the intricate $g$-dependence in (\ref{b6}) the pair cycle expression that 
contains $M_s$ is more complicated:
\begin{eqnarray}
H_s^M & = & {\rm Tr} \left \{ - \ln \left( 1 - \frac{1}{4} R_s^U U_s^{\dag} L_s^U U_s
\right)_s \right.  \nonumber \\ & & - \frac{1}{2} \left[ 1 - \frac{1}{4} R_s^U
U_s^{\dag} L_s^U U_s \right]_s^{-1} \left( (R_s^U)^2 U_s^{\dag} L_s^U U_s + R_s^U
U_s^{\dag} (L_s^U)^2 U_s \right) \nonumber \\ & & + \frac{1}{4} \left[ 1 - \frac{1}{4}
R_s^U U_s^{\dag} L_s^U U_s \right]_s^{-1} \left( (R_s^U)^3 U_s^{\dag} L_s^U U_s +
(R_s^U)^2 U_s^{\dag} (L_s^U)^2 U_s + R_s^U U_s^{\dag} (L_s^U)^3 U_s \right) \nonumber
\\ & & + \left. \frac{1}{2} \left( \frac{1}{4} \left[ 1 - \frac{1}{4} R_s^U U_s^{\dag}
L_s^U U_s \right]_s^{-1} \left( (R_s^U)^2 U_s^{\dag} L_s^U U_s + R_s^U U_s^{\dag}
(L_s^U)^2 U_s \right) \right)^2 \right \} .
\label{b14}
\end{eqnarray}
Here we used the identity
\begin{equation}
{d \over d g} B = \frac{1}{g} \left( B^2 - B \right) ,
\label{b15}
\end{equation}
where $B=[1-g A]^{-1}$ for a 4-point function, $A$. Using the identities in (\ref{ap9})
and (\ref{ap14}), one can show that 
\begin{eqnarray}
H_s^M & = & {\rm Tr} \left \{ - \ln \left( 1 - \frac{1}{4} C_C R_s C_C^{\dag} L_s
\right)_s + \frac{1}{8} C_C C_R C_C^{\dag} L_s + \frac{1}{8} C_C R_s C_C^{\dag} C_L +
\frac{1}{4} C_C^{\dag} C_C \right.  \nonumber \\ & & - \left.  \frac{1}{4} C_C
C_C^{\dag} L_s - \frac{1}{4} C_C R_s C_C^{\dag} - \frac{1}{32} C_C C_C^{\dag} L_s C_C
C_C^{\dag} L_s - \frac{1}{32} C_C R_s C_C^{\dag} C_C R_s C_C^{\dag} \right \} .
\label{b16}
\end{eqnarray}
Now one can add all the $H$'s for the {\it ss}-fermion pairs:
\begin{eqnarray}
H_{ss}& = & H_s^M + H_s^R+ H_s^L \nonumber \\ & = & {\rm Tr} \left \{ \ln \left(
1 - \frac{1}{4} C_C R_s C_C^{\dag} L_s \right)_s + \ln \left( 1 + \frac{1}{2} C_R
\right)_s + \ln \left( 1 + \frac{1}{2} C_L \right)_s \right.  \nonumber \\ & & -
\left.  \frac{1}{2} C_R - \frac{1}{2} C_L + \frac{1}{8} C_R^2 + \frac{1}{8}
C_L^2 + \frac{1}{4} C_C C_C^{\dag} \right \} .
\label{b17}
\end{eqnarray}
The resulting expression is relatively simple thanks to extensive cancellations
among the different $H$'s.

The remaining two types of fermion pairs follow exactly the same procedure and the
expressions that are obtained are also very similar to these. We quote all the
relevant expressions without unnecessary discussion.

\subsubsection{{\it ds}-fermion pairs}

The two simple 4-point functions that can directly form pair cycles in the {\it
ds}-fermion pairs are $U_d^R$ and $U_d^L$ and the simple 4-point function group that
can form pair cycles is
\begin{equation} 
M_{ds}(g) = g^2 [1-g U^R_d]_d^{-1} U_d^{\prime} [1-g U^L_d]_d^{-1} U_d^{\prime} =
g^2 R_d^U U_d^{\prime} L_d^U U_d^{\prime} .
 \label{b18}
\end{equation}

The pair cycles are formed by
\begin{equation} 
A_d(g) = \frac{1}{2} {\rm Tr} \left \{ \sum_{n=1}^{\infty} {1\over n} M(g)^n \right \}
= - \frac{1}{2} {\rm Tr} \left \{ \ln \left( 1 - M(g) \right)_d \right
\}
\label{b19} 
\end{equation} 
where $M(g)$ denotes $M_{ds}(g)$, $g U_d^R$ or $g U_d^L$.  There is a symmetry with
respect to a front-to-back transformation of these pair cycles.  The factor of
$\frac{1}{2}$ is necessary to remove the overcounting which occurs as a result of this
symmetry.  For $U_d^R$ and $U_d^L$ we have the following expressions  for (\ref{b5}),
using (\ref{b19}):
\begin{eqnarray}
H_d^R & = &{\rm Tr} \left \{ - \frac{1}{2} \ln \left( 1 - U_d^R \right)_d -
\frac{1}{2} \left[ 1 - U_d^R \right]_d^{-1} U_d^R \right.  \nonumber \\ & & + \left. 
\frac{1}{4} \left[ 1 - U_d^R \right]_d^{-1} U_d^R \left[ 1 - U_d^R \right]_d^{-1}
U_d^R \right \}
\label{b20}
\end{eqnarray}
and
\begin{eqnarray}
H_d^L & = &{\rm Tr} \left \{ - \frac{1}{2} \ln \left( 1 - U_d^L \right)_d -
\frac{1}{2} \left[ 1 - U_d^L \right]_d^{-1} U_d^L \right.  \nonumber \\ & & + \left. 
\frac{1}{4} \left[ 1 - U_d^L \right]_d^{-1} U_d^L \left[ 1 - U_d^L \right]_d^{-1}
U_d^L \right \} .
\label{b21}
\end{eqnarray}
One can show, from (\ref{ap15ru}) and (\ref{ap15lu}), that
\begin{equation}
\left[ 1 - U_d^R \right]_d^{-1} U_d^R = C_R - C_0 L C_0
\label{b22}
\end{equation}
and
\begin{equation}
\left[ 1 - U_d^L \right]_d^{-1} U_d^L = C_L - C_0 R C_0 .
\label{b23}
\end{equation}
Using (\ref{b22}) and (\ref{ap16}) one finds that (\ref{b20}) becomes
\begin{eqnarray}
H_d^R & = & {\rm Tr} \left \{ \frac{1}{2} \ln \left( 1 - C_0 L C_0 R \right)_d +
\frac{1}{2} \ln \left( 1 + C_R \right)_d - \frac{1}{2} \left( C_R - C_0 L C_0 \right)
\right.  \nonumber \\ & & + \left.  \frac{1}{4} \left(C_R - C_0 L C_0 \right)^2 \right
\} .
\label{b24}
\end{eqnarray}
Similarly, from (\ref{b23}) and (\ref{ap17}), (\ref{b21}) becomes
\begin{eqnarray}
H_d^L & = & {\rm Tr} \left \{ \frac{1}{2} \ln \left( 1 - C_0 R C_0 L \right)_d +
\frac{1}{2} \ln \left( 1 + C_L \right)_d - \frac{1}{2} \left( C_L - C_0 R C_0 \right)
\right.  \nonumber \\ & & + \left.  \frac{1}{4} \left(C_L - C_0 R C_0 \right)^2 \right
\} .
\label{b25}
\end{eqnarray}
The pair cycle expression that contains $M_{ds}$ is:
\begin{eqnarray}
H_{ds}^M & = & {\rm Tr} \left \{ - \frac{1}{2} \ln \left( 1 - R_d^U U_d^{\prime}
L_d^U U_d^{\prime} \right)_d \right.  \nonumber \\ & & - \left[ 1 - R_d^U
U_d^{\prime} L_d^U U_d^{\prime} \right]_d^{-1} \left( (R_d^U)^2 U_d^{\prime} L_d^U
U_d^{\prime} + R_d^U U_d^{\prime} (L_d^U)^2 U_d^{\prime} \right) \nonumber \\ &
& + \frac{1}{2} \left[ 1 - R_d^U U_d^{\prime} L_d^U U_d^{\prime} \right]_d^{-1}
\left( (R_d^U)^3 U_d^{\prime} L_d^U U_d^{\prime} + (R_d^U)^2 U_d^{\prime}
(L_d^U)^2 U_d^{\prime} + R_d^U U_d^{\prime} (L_d^U)^3 U_d^{\prime} \right)
\nonumber \\ & & + \left.  \frac{1}{4} \left( \left[ 1 - R_d^U U_d^{\prime}
L_d^U U_d^{\prime} \right]_d^{-1} \left( (R_d^U)^2 U_d^{\prime} L_d^U U_d^{\prime}
+ R_d^U U_d^{\prime} (L_d^U)^2 U_d^{\prime} \right) \right)^2 \right \} ,
\label{b26}
\end{eqnarray}
where we again used the identity in (\ref{b15}). The identities in 
(\ref{ap15}) lead to
\begin{eqnarray}
H_d^M & = & {\rm Tr} \left \{ - \frac{1}{2} \ln \left( 1 - C_0 L C_0 R \right)_d +
\frac{1}{2} C_0 C_R C_0 L + \frac{1}{2} C_0 R C_0 C_L + \frac{1}{2} C_0^2 \right. 
\nonumber \\ & & - \left.  \frac{1}{2} C_0 C_0 L - \frac{1}{2} C_0 R C_0 - \frac{1}{4}
C_0 C_0 L C_0 C_0 L - \frac{1}{4} C_0 R C_0 C_0 R C_0 \right \} .
\label{b27}
\end{eqnarray}
Now one can add all the $H$'s for the {\it sd}-fermion pairs:
\begin{eqnarray}
H_{ds}& = & H_{ds}^M + H_d^R+ H_d^L \nonumber \\ & = & {\rm Tr} \left \{ \frac{1}{2}
\ln \left( 1 - C_0 L C_0 R \right)_d + \frac{1}{2} \ln \left( 1 + C_R \right)_d +
\frac{1}{2} \ln \left( 1 + C_L \right)_d \right.  \nonumber \\ & & - \left. 
\frac{1}{2} C_R - \frac{1}{2} C_L + \frac{1}{4} C_R^2 + \frac{1}{4} C_L^2 +
\frac{1}{2} C_0^2
\right \} .
\label{b28}
\end{eqnarray}
Here too we have a relatively simple expression as a result of cancellations among the
different $H$'s.

\subsubsection{{\it dd}-fermion pairs}

There is only one simple 4-point function, $U_d^0$, which can directly form a pair
cycle in the {\it dd}-fermion pairs. The simple 4-point function group for this type
of fermion pairs is
\begin{equation} 
M_{dd}(g) = g^2 [1-g U_d^0]_d^{-1} U_d^{\dag} [1- g U_d^0]_d^{-1} U_d = g^2 B
U_d^{\dag} B U_d .
\label{b29} 
\end{equation} 
Because $U_d^0$ looks different under a front-to-back transformation we shall use it
twice.  These two pair cycles are then related through a back-to-front transformation
and each must be multiplied by $\frac{1}{2}$.  The pair cycle for the simple 4-point
function group of (\ref{b29}) has a front-to-back symmetry.  The pair cycles are
therefore formed by the expression in (\ref{b19}) where $M(g)$ denotes
$M_{dd}(g)$ or $g U_d^0$.  

Substituting $U_d^0$ into (\ref{b5}), using (\ref{b19}), gives
\begin{eqnarray}
H_d^0 & = &{\rm Tr} \left \{ - \frac{1}{2} \ln \left( 1 - U_d^0 \right)_d -
\frac{1}{2} \left[ 1 - U_d^0 \right]_d^{-1} U_d^0 \right.  \nonumber \\ & & +
\left.  \frac{1}{4} \left[ 1 - U_d^0 \right]_d^{-1} U_d^0 \left[ 1 - U_d^0
\right]_d^{-1} U_d^0 \right \} .
\label{b30}
\end{eqnarray}
From this expression one can use two different sets of expressions to get two different
results which are related by a back-to-front transformation. First one can use
(\ref{ap20cc}) and (\ref{ap20z}), to show that
\begin{equation}
\left[ 1 - U_d^0 \right]_d^{-1} U_d^0 = C_0 - C_C^{\dag} Z C_C
\label{b31}
\end{equation}
or one can use (\ref{ap20c}) and (\ref{ap20zc}), to show that
\begin{equation}
\left[ 1 - U_d^0 \right]_d^{-1} U_d^0 = C_0 - C_C Z C_C^{\dag}
\label{b32} .
\end{equation}
Substituting (\ref{b31}) and (\ref{ap22}) into (\ref{b30}) one 
finds
\begin{eqnarray} 
H_d^0 & = & {\rm Tr} \left \{ \frac{1}{2} \ln \left(1 - C_C^{\dag} Z C_C Z \right)_d
+ \frac{1}{2} \ln \left(1 + C_0 \right)_d - \frac{1}{2} \left( C_0 - C_C^{\dag} Z
C_C \right) \right.  \nonumber \\ & & + \left.  \frac{1}{4} \left( C_0 - C_C^{\dag}
Z C_C \right)^2 \right \}
\label{b33} 
\end{eqnarray} 
and substituting (\ref{b32}) and (\ref{ap21}) into (\ref{b30}) one 
finds
\begin{eqnarray} 
H_d^{0\dag} & = & {\rm Tr} \left \{ \frac{1}{2} \ln \left(1 - C_C Z C_C^{\dag} Z
\right)_d + \frac{1}{2} \ln \left(1 + C_0 \right)_d - \frac{1}{2} \left( C_0 - C_C
Z C_C^{\dag} \right) \right.  \nonumber \\ & & + \left.  \frac{1}{4} \left( C_0 -
C_C Z C_C^{\dag} \right)^2 \right \} ,
\label{b34}
\end{eqnarray} 
where we placed a $\dag$ on the $H$ in the latter expression to show that the 
chiralities on the fermion lines are interchanged. The pair cycle terms 
that contain $M_{ds}$ are:
\begin{eqnarray}
H_{dd}^M & = & {\rm Tr} \left \{ - \frac{1}{2} \ln \left( 1 - B U_d^{\dag} B U_d
\right)_d - \left[ 1 - B U_d^{\dag} B U_d \right]_d^{-1} \left( B^2 U_d^{\dag} B U_d
+ B U_d^{\dag} B^2 U_d \right) \right.  \nonumber \\ & & + \frac{1}{2} \left[ 1
- B U_d^{\dag} B U_d \right]_d^{-1} \left( B^3 U_d^{\dag} B U_d + B^2 U_d^{\dag}
B^2 U_d + B U_d^{\dag} B^3 U_d \right) \nonumber \\ & & + \left.  \frac{1}{4}
\left( \left[ B U^{\dag}_d B U_d \right]_d^{-1} \left( B^2 U_d^{\dag} B U_d + B
U_d^{\dag} B^2 U_d \right) \right)^2 \right \} ,
\label{b35}
\end{eqnarray}
where we used the identity in (\ref{b15}). The expressions in 
(\ref{ap20}) then gives
\begin{eqnarray}
H_{dd}^M & = & {\rm Tr} \left \{ - \frac{1}{2} \ln \left( 1 - C_C Z C_C^{\dag} Z
\right)_d + \frac{1}{2} C_C C_0 C_C^{\dag} Z + \frac{1}{2} C_C Z C_C^{\dag} C_0 +
\frac{1}{2} C_C C_C^{\dag} \right.  \nonumber \\ & & - \left.  \frac{1}{2} C_C
C_C^{\dag} Z - \frac{1}{2} C_C Z C_C^{\dag} - \frac{1}{4} C_C C_C^{\dag} Z C_C
C_C^{\dag} Z - \frac{1}{4} C_C Z C_C^{\dag} C_C Z C_C^{\dag} \right \} .
\label{b36}
\end{eqnarray}
Adding all the $H$'s for the {\it dd}-fermion pairs, one finds
\begin{eqnarray}
H_{dd}& = & H_{dd}^M + H_d^0+ H_d^{0\dag} \nonumber \\ & = & {\rm Tr} \left \{
\frac{1}{2} \ln \left( 1 - C_C Z C_C^{\dag} Z \right)_d + \ln \left( 1 + C_0 \right)_d
- C_0 + \frac{1}{2} C_0^2 + \frac{1}{2} C_C C_C^{\dag} \right \} .
\label{b37}
\end{eqnarray}

\subsection{The effective action}

Now one can construct the expression for the pair cycle terms of the effective action
(\ref{b0}) by adding the expressions in (\ref{b4}), (\ref{b17}), (\ref{b28}) and
(\ref{b37})
\begin{eqnarray}
H_{Total}& = & H_{sd} + H_{ss} + H_{ds} + H_{dd} \nonumber \\ & = & {\rm Tr}
\left \{ \ln \left(1 - \frac{1}{4} R_s C_C^{\dag} L_s C_C \right)_s + \ln \left(1 +
\frac{1}{2} C_R \right)_s + \ln \left(1 + \frac{1}{2} C_L \right)_s \right.
\nonumber \\ & & + \frac{1}{2} \ln \left(1 - R C_0 L C_0 \right)_d + \frac{1}{2}
\ln \left(1 + C_R \right)_d + \frac{1}{2} \ln \left(1 + C_L \right)_d + \ln
\left(1 + C_0 \right)_d \nonumber \\ & & + \frac{1}{2} \ln \left(1 - Z C_C^{\dag}
Z C_C \right)_d + \ln(1+C_0)_s - C_R - C_L - 2 C_0 \nonumber \\ & & \left.+
\frac{3}{4} C_C C_C^{\dag} + \frac{3}{2} C_0^2 + \frac{3}{8} C_R^2 + \frac{3}{8}
C_L^2 \right \} .
\label{b38}
\end{eqnarray}

One can then combine the different parts in (\ref{e3}), (\ref{e4}), (\ref{e5}) and
(\ref{b38}) to construct the generating functional, $W$ -- i.e.\ the set of all
connected vacuum diagrams.  Next one can perform the Legendre transform, (\ref{ei7}),
which removes (\ref{e5}) to produce the expression of the full effective action.  With
all the explicit $J$-dependence removed, the effective action becomes a functional of
the 4-point functions:
\begin{eqnarray}
\Gamma & = & W_c - W_p + W_{pp} - W_{art} + W_{skel} \nonumber \\ & = & {\rm Tr}
\left \{ \ln \left(1 - \frac{1}{4} R_s C^{\dag} L_s C \right)_s + \ln \left(1 +
\frac{1}{2} C_R \right)_s + \ln \left(1 + \frac{1}{2} C_L \right)_s +
\ln(1+C_0)_s \right.  \nonumber \\ & & + \frac{1}{2} \ln \left(1 - R C_0 L C_0
\right)_d + \frac{1}{2} \ln \left(1 + C_R \right)_d + \frac{1}{2} \ln \left(1 +
C_L \right)_d + \ln \left(1 + C_0 \right)_d \nonumber \\ & & + \frac{1}{2} \ln
\left(1 - Z C^{\dag} Z C \right)_d - C_R - C_L - 2 C_0 \nonumber \\ & & \left.+
\frac{1}{2} C C^{\dag} + C_0^2 + \frac{1}{4} C_R^2 + \frac{1}{4} C_L^2 \right \}
+ V_{4PI}[C,C^{\dag},C_0,C_R,C_L] .
\label{e7}
\end{eqnarray}
This result for the effective action is the main result of our analysis. It can be applied to
arbitrary models, and it provides the means to determine which
solution of the gap equations of Sec.~\ref{gap} represents the
true vacuum of the theory.

Here we want to remark on the large $N_c$ expansion of this effective action, which is
discussed in Sec.~\ref{over}. When such an expansion is made and only the leading
terms are retained, one notes that $H_{ss}$ in (\ref{b17}) and $H_{sd}$ in (\ref{b4})
are suppressed because their fermion loops cannot form color loops. Next one notes that
$H_{ds}$ in (\ref{b28}) and $H_{dd}$ in (\ref{b37}) decouple from each other because
they do not share the same $C$'s. (Each of $H_{ds}$ and $H_{dd}$ contain one of the
respective color structures of $C_0$ -- see discussion below (\ref{o14}).) This leads
to the two decoupled effective actions which are presented in (\ref{nc1}) and
(\ref{nc2}).

\section{Obtaining gap equations from the effective action}
\label{derive}

The gap equations for the 4-point functions (\ref{v6}) are obtained by taking
the functional derivatives of the effective action (\ref{e7}), as in (\ref{ei12}), and then
amputating the result. A key step in this process is to take the functional derivatives of the
pair cycle terms which are provided in (\ref{b4}), (\ref{b17}), (\ref{b28}) and (\ref{b37}). It
is necessary to take special care of the assignments of external lines which result from these
functional derivatives. For the case of $C_0$ this is just
\begin{equation}
{\delta C_0(1,2,3,4) \over \delta C_0(a,b,c,d)} = \delta_a^1 \delta_b^2 
\delta_c^3 \delta_d^4 ,
\label{b40}
\end{equation}
where we use $1,2,3,4$ and $a,b,c,d$ to distinguish external fermion lines. For 
the other four cases, $C_C$, $C_C^{\dag}$, $C_R$ and $C_L$, we have
\begin{equation}
{\delta C(1,2,3,4) \over \delta C(a,b,c,d)} = [\delta_a^1 \delta_c^3 + 
\delta_c^1 \delta_a^3 ] [ \delta_b^2 \delta_d^4 + \delta_d^2 \delta_b^4 ] .
\label{b41}
\end{equation}
The latter functional derivative gives rise to four terms. Often these four terms may
be related to each other through some interchanges of external lines. These
transformations are defined in (\ref{v4}). 

Next we calculate the functional derivatives of the $H$'s in (\ref{b4}), (\ref{b17}),
(\ref{b28}) and (\ref{b37}) with respect to $C_0$, $C_C^{\dag}$ and $C_R$. Those of
$C_C$ and $C_L$ are related to these in an obvious way. In each case we shall find
that the functional derivative of the pair cycles of a specific type of fermion pairs
with respect to a specific 4-point function reproduces a specific $T$ (or more than
one if they are related by one of the above transformations). 

We start with $C_0$. Only the pair cycles with the {\it sd}-, {\it ds}- and {\it
dd}-fermion pairs depend on $C_0$. Their functional derivatives are
\begin{equation}
\left. {\delta H_{sd} \over \delta C_0} \right|_{amp} =  \left[ 1 + C_0 \right]_s^{-1}
- 1 + C_0 = T_s^0 ;
\label{b42}
\end{equation}
\begin{eqnarray}
\left. {\delta H_{ds} \over \delta C_0} \right|_{amp} & = & \frac{1}{2} \left[ 1 - R C_0 L
C_0 \right]_d^{-1} \left( - R C_0 L \right) + \frac{1}{2} \left( - L C_0 \left[ 1 - R C_0 L
C_0 \right]_d^{-1} R \right)^{RF} + C_0 \nonumber \\ & = & C_0 - R C_0 L \left[ 1 - C_0
R C_0 L \right]_d^{-1} = T_d^{\prime} ;
\label{b43}
\end{eqnarray}
\begin{eqnarray}
\left. {\delta H_{dd} \over \delta C_0} \right|_{amp} & = & \frac{1}{2} \left( Z \left[ 1 -
C_C Z C_C^{\dag} Z \right]_d^{-1} C_C Z C_C^{\dag} Z \right) + \frac{1}{2} \left( Z
\left[ 1 - C_C^{\dag} Z C_C Z \right]_d^{-1} C_C^{\dag} Z C_C Z \right)^{RF}
\nonumber \\ & & + \left[ 1 + C_0 \right]_d^{-1} - 1 + C_0 \nonumber \\ & = & C_0 + Z
\left[ 1 - C_C Z C_C^{\dag} Z \right]_d^{-1} - 1 = T_d^0 .
\label{b44}
\end{eqnarray}
The $RF$ transformations shown in (\ref{b43}) and (\ref{b44}) are equal to the $RF$
transformations applied to each individual object in reversed order in each of the
terms. This reproduces the three $T$'s that appear in the gap equation for $C_0$,
(\ref{ap5}), (\ref{ap26a}) and (\ref{ap27d}).  Next we consider $C_C^{\dag}$.  Only
the pair cycles in the {\it ss}- and {\it dd}-fermion pairs depend on $C_C^{\dag}$:
\begin{eqnarray}
\left. {\delta H_{ss} \over \delta C_C^{\dag}} \right|_{amp} & = & \left[ 1 - \frac{1}{4}
L_s C_C R_s C_C^{\dag} \right]_s^{-1} \left( -  \frac{1}{4} L_s C_C R_s \right) \left(
\times 4 \right) + \frac{1}{4} C_C \left( \times 4 \right) \nonumber \\ & = & C_C - L_s C_C
R_s \left[ 1 - \frac{1}{4} C_C^{\dag} L_s C_C R_s \right]_s^{-1} = T_s ;
\label{b45}
\end{eqnarray}
\begin{eqnarray}
\left. {\delta H_{dd} \over \delta C_C^{\dag}} \right|_{amp} & = & \frac{1}{2} \left[ 1 - Z
C_C Z C_C^{\dag} \right]_d^{-1} \left( - Z C_C Z \right) \left( 1 + 1^{RF} + 1^T +
1^{RFT} \right) \nonumber \\ & & + \frac{1}{2} C \left( 1 + 1^{RF} + 1^T + 1^{RFT}
\right) \nonumber \\ & = & \left( C_C - Z C_C Z \left[ 1 - C_C^{\dag} Z C_C Z
\right]_d^{-1} \right) \left( 1 + 1^T \right) = T_d + \left( T_d \right)^T .
\label{b46}
\end{eqnarray}
The $(\times 4)$ comes from the {\it ss}-symmetry which generates four identical terms.
The $(1 + 1^{RF} + 1^T + 1^{RFT})$ denotes different exchanges of external lines.
Because the diagram is symmetric under the combined $RF$-transformation, the
expression reduces to the final expression with a factor of 2. Thus we reproduce the
three $T$'s which appear in the gap equation for $C_C$, (\ref{ap25a}) and
(\ref{ap27a}). The gap equation for $C_C^{\dag}$ is quite similar. One merely
interchanges the chiralities to get the necessary expressions. For $C_R$ only the {\it
ss}- and {\it ds}-fermion pairs need to be considered:
\begin{eqnarray}
\left. {\delta H_{ss} \over \delta C_R} \right|_{amp} & = & \left( - \frac{1}{2} R_s \left[ 1 -
\frac{1}{4} C_C^{\dag} L_s C_C R_s \right]_s^{-1} \left(- \frac{1}{4} C_C^{\dag} L_s
C_C R_s \right) + \frac{1}{2} R_s - \frac{1}{2} + \frac{1}{4} C_R \right) \left( \times 4
\right) \nonumber \\ & = & C_R + 2 R_s \left[ 1 - \frac{1}{4} C_C^{\dag} L_s C_C R_s
\right]_s^{-1} - 2 = T_s^R ;
\label{b47}
\end{eqnarray}
\begin{eqnarray}
\left. {\delta H_{ds} \over \delta C_R} \right|_{amp} & = & \left( - \frac{1}{2} R \left[ 1 -
C_0 L C_0 R \right]_d^{-1} \left( - C_0 L C_0 R \right) + \frac{1}{2} R \right. \nonumber \\
& & - \left.  \frac{1}{2} + \frac{1}{2} C_R \right) \left( 1 + 1^{RF} + 1^T + 1^{RFT}
\right) \nonumber \\ & = & \left( C_R + R \left[ 1 - C_0 L C_0 R \right]_d^{-1} - 1
\right) \left( 1 + 1^T \right) = T_d^R + \left( T_d^R \right)^T .
\label{b48}
\end{eqnarray}
So finally we reproduced the three $T$'s which appear in the $C_R$ gap
equation -- (\ref{ap25c}) and (\ref{ap26c}).  Those for $C_L$ would be identical apart
from replacing $R$ with $L$.

Next we find the functional derivatives of $W_{art}$ with respect to the various
4-point functions. (Amputation is implied in all of the following.)
\begin{equation}
{\delta W_{art} \over \delta C_C^{\dag}} = C_C , ~~~
{\delta W_{art} \over \delta C_C} = C_C^{\dag} , ~~~
{\delta W_{art} \over \delta C_0} = C_0 , ~~~
{\delta W_{art} \over \delta C_R} = C_R ~~~ {\rm and} ~~~
{\delta W_{art} \over \delta C_L} = C_L .
\label{e12}
\end{equation}
Finally the functional derivatives of $W_{skel}$ with respect to the various 4-point
functions gives rise to the 4PI 4-point diagrams which appear as the $K$-terms in the
gap equations:
\begin{equation}
{\delta W_{skel} \over \delta C_C^{\dag}} = K_C , ~~~
{\delta W_{skel} \over \delta C_C} = K_C^{\dag} , ~~~
{\delta W_{skel} \over \delta C_0} = K_0 , ~~~
{\delta W_{skel} \over \delta C_R} = K_R ~~~ {\rm and} ~~~
{\delta W_{skel} \over \delta C_L} = K_L .
\label{e13}
\end{equation}
Collecting the different terms for each specific 4-point function according to
(\ref{e2}), using (\ref{b0}) and the first line of (\ref{b38}), one reproduces the gap
equations exactly as in~(\ref{v6}).

\end{document}